%% file: main.tex
\documentclass[lettersize,journal]{IEEEtran}
\usepackage{amsmath,amsfonts}
\usepackage{algorithmic}
\usepackage{algorithm}
\usepackage{array}
\usepackage[caption=false,font=normalsize,labelfont=sf,textfont=sf]{subfig}
\usepackage{textcomp}
\usepackage{stfloats}
\usepackage{url}
\usepackage{verbatim}
\usepackage{graphicx}
\usepackage{xcolor}
\usepackage[english]{babel}
\usepackage[autostyle]{csquotes}
\usepackage{cite}
\usepackage{mathtools}
\usepackage{adjustbox}
\usepackage{caption}

\newlength\myindent
\input{macros}

\usepackage{enumitem}

\usepackage{multirow}
\def\BibTeX{{\rm B\kern-.05em{\sc i\kern-.025em b}\kern-.08em
    T\kern-.1667em\lower.7ex\hbox{E}\kern-.125emX}}
\usepackage{float}
\usepackage{caption}

\newcommand\algorithmicprocedure{\textbf{procedure}}
\newcommand{\algorithmicendprocedure}{\algorithmicend\ \algorithmicprocedure}
\makeatletter
\newcommand\PROCEDURE[3][default]{%
  \ALC@it
  \algorithmicprocedure\ \textsc{#2}(#3)%
  \ALC@com{#1}%
  \begin{ALC@prc}%
}
\newcommand\ENDPROCEDURE{%
  \end{ALC@prc}%
  \ifthenelse{\boolean{ALC@noend}}{}{%
    \ALC@it\algorithmicendprocedure
  }%
}
\newenvironment{ALC@prc}{\begin{ALC@g}}{\end{ALC@g}}
\makeatother

\makeatletter
\renewcommand{\fnum@figure}{Fig. \thefigure}
\makeatother

\usepackage{amssymb}
\usepackage{nomencl}

\makenomenclature
\usepackage{etoolbox}
\renewcommand\nomgroup[1]{%
  \item[\bfseries
  \ifstrequal{#1}{A}{Abbreviations}{
  \ifstrequal{#1}{S}{Sets}{
  \ifstrequal{#1}{P}{Parameters}{%
  \ifstrequal{#1}{V}{Variables}{%
  \ifstrequal{#1}{I}{Indices}{}}}}}%
]}

\begin{document}

\title{Charge Manipulation Attacks Against Smart Electric Vehicle Charging Stations and Deep Learning-based Detection Mechanisms}


\author{Hamidreza Jahangir, Subhash Lakshminarayana~\IEEEmembership{Senior Member, IEEE}, and H. Vincent Poor~\IEEEmembership{Life Fellow, IEEE}
\thanks{H. Jahangir and S. Lakshminarayana (Corresponding author) are with the School of Engineering, University of Warwick, CV47AL, UK.
H. V. Poor is with the Department of Electrical and Computer Engineering, Princeton University, Princeton, NJ 08544, USA. E-mails: (Hamidreza.Jahangir@ieee.org, Subhash.Lakshminarayana@warwick.ac.uk, and poor@princeton.edu).}}

\markboth{ }%
{}


\maketitle

\begin{abstract}
The widespread deployment of ``smart'' electric vehicle charging stations (EVCSs) will be a key step toward achieving green transportation. The connectivity features of smart EVCSs can be utilized to schedule EV charging operations while respecting user preferences, thus avoiding synchronous charging from a large number of customers and relieving grid congestion. 
However, the communication and connectivity requirements involved in smart charging raise cybersecurity concerns.  
In this work, we investigate \emph{charge manipulation attacks} (CMAs) against EV charging, in which an attacker manipulates the information exchanged during smart charging operations. The objective of CMAs is to shift the  EV aggregator's demand across different times of the day. The proposed CMAs can bypass existing protection mechanisms in EV communication protocols. We quantify the impact of CMAs on the EV aggregator's economic profit by modeling their participation in the day-ahead (DA) and real-time (RT) electricity markets. Finally, we propose 
an unsupervised deep learning-based mechanism to detect CMAs by monitoring the parameters involved in EV charging. We extensively analyze the attack impact and the efficiency of the proposed detection on real-world EV charging datasets. The results highlight the vulnerabilities of smart charging operations and the need for a monitoring mechanism to detect malicious CMAs.
\end{abstract}





\nomenclature[S]{\(\mathcal{N}\)}{\scriptsize {Set of EVCPs, indexed by \textit{n}}}
\nomenclature[S]{\(\mathcal{N}_h\)}{\scriptsize {Set of hacked EVCPs, indexed by $n_{hd}$, $n_{he}$, $n_{hs}$ }}
\nomenclature[S]{\(\mathcal{N}_{hs}\)}{\scriptsize {Set of hacked EVCPs with manipulated start time, indexed by $n_{hs}$}}
\nomenclature[S]{\(\mathcal{N}_{he}\)}{\scriptsize {Set of hacked EVCPs with manipulated end time, indexed by $n_{he}$}}
\nomenclature[S]{\(\mathcal{N}_{hd}\)}{\scriptsize {Set of hacked EVCPs with manipulated demand, indexed by $n_{hd}$}}
\nomenclature[S]{\(\mathcal{T}_a\)}{\scriptsize {Set of charging intervals ($\Delta$t-minute steps), indexed by $t[1,288]$ (for launching CMAs)}}

\nomenclature[S]{\(\mathcal{T}\)}{\scriptsize {Set of bidding time intervals (hour), indexed by $tm[1,24]$~(for energy markets)}}
\nomenclature[S]{\(\hat{\mathcal{T}}\)}{\scriptsize {Set of five-minute market time intervals, indexed by $\hat{tm}[1,12]$~(for energy markets)}}

\nomenclature[V]{\(Ch_{t}^{n}, \overline{Ch}_{t+\Delta t}^{n}\)}{\scriptsize {Charging rate EVCP $n$ time $t$, $t+\Delta t$ (kW)}}
\nomenclature[V]{\(Av_{t}^{n}, \overline{Av}_{t+\Delta t}^{n} \)}{\scriptsize {Available time intervals EVCP $n$ time $t$, $t+\Delta t$ }}
\nomenclature[V]{\(et_{t}^{n}, \overline {et}_{t+\Delta t}^{n}\)}{\scriptsize {Requested end time EVCP $n$ time $t$, $t+\Delta t$} ($\Delta t$-min)}
\nomenclature[V]{\(st_{t}^{n}, \overline {st}_{t+\Delta t}^{n}\)}{\scriptsize {Requested start time EVCP $n$ time $t$, $t+\Delta t$ ($\Delta t$-min)}}
\nomenclature[V]{\(d_{t}^{n}, \overline {d}_{t+\Delta t}^{n}\)}{\scriptsize {Requested demand EVCP $n$ time $t$, $t+\Delta t$ (kWh)}}
\nomenclature[V]{\({Ta}_{t}^{n}, \overline {Ta}_{t+\Delta t}^{n}\)}{\scriptsize {Total available time EVCP $n$ time $t$, $t+\Delta t$}}
\nomenclature[V]{\(y,\hat{y}\)}{\scriptsize {Real value, Predicted value}}

\nomenclature[V]{\(Cost^{CH,DA}\)}{\scriptsize {DA charging cost(\$)}}
\nomenclature[V]{\(Cost^{DA}/Cost^{RT}\)}{\scriptsize {Total DA/RT aggregator cost(\$)}}
\nomenclature[V]{\(Cost^{EENC,DA}/Cost^{EENC,RT}\)}{\scriptsize {DA/RT EENC cost(\$)}}
\nomenclature[V]{\(Cost^{INC,RT}\)}{\scriptsize {Cost of placing incremental bids in RTM(\$)}}
\nomenclature[V]{\(Cost^{PEN,RT}\)}{\scriptsize {Penalty cost of not honoring DAM bids in RTM(\$)}}
\nomenclature[V]{\(Cost^{Total}\)}{\scriptsize {Total aggregator cost(\$)}}
\nomenclature[V]{\(ENS^{DA}_n/ENS^{RT}_n \)}{\scriptsize {DA/RT energy not supplied of EVCP $n$~(kWh)}}
\nomenclature[V]{\(P^{INC,RT}_{t_m,\hat{t_m}} \)}{\scriptsize {Incremental energy bids in RTM in hour  $t_m$ and time interval $\hat{t_m}$~(kW)}}
\nomenclature[V]{\(P^{PEN, RT}_{t_m,\hat{t_m}} \)}{\scriptsize {Amount of DAM awarded bid not consumed in RTM for hour $t_m$ and time interval $\hat{t_m}$~(kW)}}
\nomenclature[V]{\(pch^{DA}_{n,t_m,\hat{t_m}} / pch^{RT}_{n,t_m,\hat{t_m}} \)}{\scriptsize {DA/RT demand of EVCP $n$ in hour $t_m$ and time interval $\hat{t_m}$~(kW)}}
\nomenclature[V]{\(PCH^{DA}_{t_m} \)}{\scriptsize {DA demand of all EVCSs $n$ in hour $t_m$ ~(kW)}}
\nomenclature[V]{\(PCH^{RT}_{t_m,\hat{t_m}} \)}{\scriptsize {RT demand of all EVCSs $n$ in hour $t_m$ and time interval $\hat{t_m}$~(kW)}}
\nomenclature[V]{}{}
\nomenclature[V]{}{}

\nomenclature[P]{\(Pl\)}{\scriptsize {Planning horizon (h)}}
\nomenclature[P]{\(\Delta t\)}{\scriptsize {Length of time intervals (min), for launching charge manipulation attacks CMAs}}
\nomenclature[P]{\(n_{step}\)}{\scriptsize {Number of steps in the charging task}}
\nomenclature[P]{\(ACR_{att}\)}{\scriptsize {Added charging rate value to each EV under attack (kW)}}
\nomenclature[P]{\(Ch_{av}\)}{\scriptsize {Estimated average charging rate of the target EVCPs (kW)}}
\nomenclature[P]{\(C_{att}\)}{\footnotesize {Attack rate coefficient}}
\nomenclature[P]{\(T_r\)}{\scriptsize {Threshold value for anomaly detection}}

\nomenclature[P]{\(\rho_{tm}^{DA}\)}{\scriptsize {DAM price for hour $t_m$ (\$/kWh)}}
\nomenclature[P]{\(\hat{\Delta{tm}}\)}{\scriptsize {Length of five-minute time interval (hour), for energy markets}}
\nomenclature[P]{\(\Delta{tm}\)}{\scriptsize {One-hour time interval, for energy markets}}
\nomenclature[P]{\(\rho^{EENC,DA} / \rho^{EENC,RT} \)}{\scriptsize {Expected energy not charged (EENC) penalty cost in day-ahead energy market (DAM) / real-ahead energy market (RTM) (\$/kWh)}}
\nomenclature[P]{\(\rho^{PEN,RT}\)}{\scriptsize {Penalty value of not honoring DAM bids in RTM(\$/kWh)}}
\nomenclature[P]{\(\rho^{RT}_{tm, \hat{tm}}\)}{\scriptsize {RTM energy price for hour $tm$ at time interval $\hat{tm}$ (\$/kWh)}}
\nomenclature[P]{\(Demand^{DA}_{n}/ Demand^{RT}_{n}\)}{\scriptsize {DA/RT demand of EVCP $n$ (kWh)}}

\printnomenclature

\section{Introduction}\label{introduction}

\subsection{Background and Motivation}

\IEEEPARstart{E}{lectrification} of transportation is crucial to in achieving the net-zero goals set by several nations worldwide. 
However, with the rapid increase in the number of electric vehicles (EVs) and the associated charging operations, unregulated EV charging demand can quickly overwhelm power grids and push them beyond their operational limits  \cite{acharya2021cyber}. Smart charging is an essential solution for regulating the  EV charging process, as it shifts the EV charging load away from high-demand periods while conforming to users' charging preferences. Countries such as the UK, USA, and European nations are in the process of introducing smart charging regulations \cite{EV_policy}. Smart charging involves the communication and coordination of different entities, such as EV owners, electric vehicle charging stations (EVCSs), central system (CS), and energy utilities. Commercial protocols, such as the \textit{Open Charge Point Protocol} (OCPP), offer a way for orchestrating the communication and, ultimately, power flow between the EVCS and CS \cite{OCPP_survey}. 

However, the connectivity features of smart EVCSs (i.e., the ability to send/receive data from users and the CS) raise cyber security concerns \cite{OCPP_survey}. Recent works have shown that the communication interfaces involved in EV charging operations are vulnerable to cyber-attacks; for instance, the OCPP protocol used for communication between the EVCSs and the CS is known to be vulnerable to man-in-the-middle (MitM) attacks \cite{alcaraz2017ocpp, OCPP_survey}.
By exploiting vulnerabilities in various components of the smart charging ecosystem, an attacker can launch charge manipulation attacks (CMAs), i.e., tamper with the charging settings or interrupt charging operations.  Such attacks may interfere with the operation of power grids, energy market, and demand response programs. Investigating probable stealthy CMAs, taking into account the current security measures in commercial EVCS protocols, and developing detection approaches are of utmost importance.

\subsection{Literature Survey}\label{survey}



There has been a growing interest in studying attacks against EVCS in the research literature. We categorize these attacks into three main groups depending on the attack impact.  



\subsubsection{Sudden surge in demand/supply}
 Manipulating the demand from a large number of EV charging operations can cause a sudden surge or drop in the power grid demand \cite{wang2019electrical}. This can accomplished in different ways. \\
(i) Obtaining remote access to EVCSs -- An attacker with remote access to EVCSs can alter their charging settings such that a large number of 
EVs begin charging simultaneously (by triggering inactive EVCSs or boosting the charging rate of active ECVSs). Such large-scale attacks can disrupt the balance between power grid supply and demand, leading to frequency instability and cascading failures \cite{girdhar2021hidden}. Moreover, the attacker can target the power grid's peak demand periods to exacerbate the attack impact \cite{wang2019electrical}. \\
(ii) SMS phishing attacks -- Large-scale surges in demand can also be initiated by a social engineering attack \cite{soykan2021disrupting}. For instance, the attacker can launch an SMS  phishing attack, sending incorrect electricity price information to EV customers, leading them to charge in a synchronous manner. \\
(iii) False data injection (FDI) and hijacking attacks -- In \cite{gumrukcu2022impact}, a hybrid method involving FDI and hijacking attacks was examined. The attack involved the manipulation of user demand by hijacking the mobile phones (used to set the charging parameters) and simultaneous manipulation of the aggregate demand constraints set by the DSO during peak demand periods, thereby disrupting the stability of the power system.

\subsubsection{Coordinated switching attacks}
In a more sophisticated attack, attackers with remote access to EVCSs can switch on and off the EVCPs multiple times at a high frequency \cite{acharya2020public, sayed2022electric}. Such attacks can cause more severe disruptions than a one-time surge attack, such as causing frequency instability (as such attacks can relocate the power grid's eigenvalues to the unstable region \cite{jahangir2023deep}), but are more challenging to execute. A recent study  \cite{kabir2021two} addresses a stealthier version by incorporating inter-area oscillations, which are low-frequency oscillations (i.e., $\leq 0.8$ Hz) in the power grid. Due to the low frequency of these switching attacks, the power grid operator cannot easily distinguish them from the natural oscillations of the power grid \cite{hammad2017class}.

\subsubsection{Manipulating the Energy Market} Different from attacks aiming to destabilize the power grid, an attacker with access to a sizable number of high-wattage Internet of Things (IoT) devices, such as EVCSs, can manipulate the energy markets and gain financial incentives \cite{shekari2021mamiot, soltan2019protecting, acharya2022false, soltan2018blackiot}. For instance, by deliberately creating surges or drops in the energy demand, the attacker can cause volatility in the energy prices and take advantage of these fluctuations to make a profit (see Section~\ref{energy_market} for further details).

\begin{figure}[]
\centering
\includegraphics[height=6.5 cm,width=9.5 cm,trim= 0 10 90 0,clip]{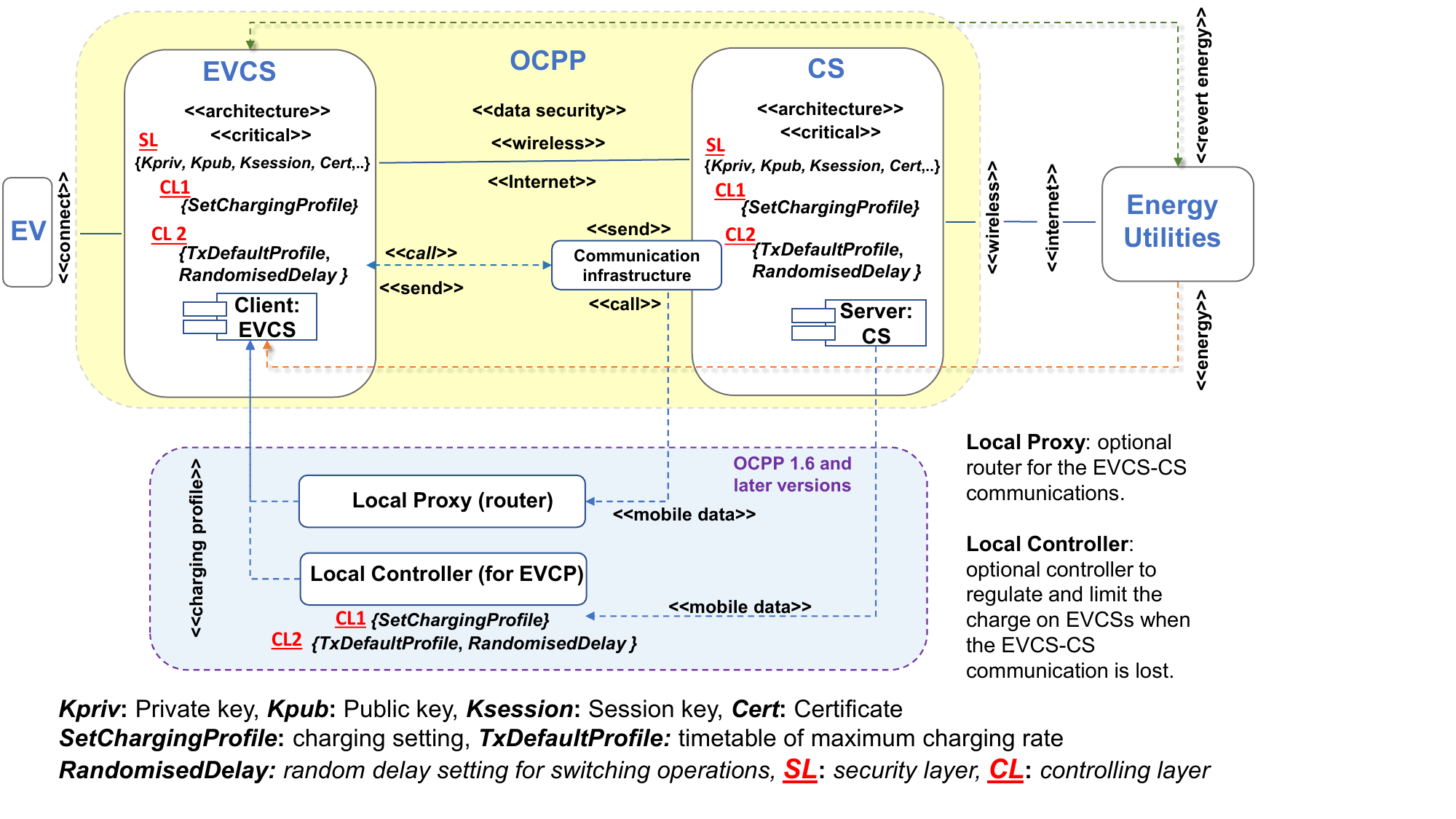}
\caption{Schematic of the OCPP protocol implementation }
\label{fig:OCPP}
\vspace{+0.2 cm}
\end{figure}

{\bf Drawbacks of Existing Works:} 
Despite the growing literature on this topic, most existing works consider only direct manipulation of EV charging operations (sudden surge in demand/supply or coordinated charging attacks) while ignoring the smart charging aspects. However, existing security measures in commercial EV protocols such as OCPP make it extremely hard to execute such direct manipulations. A schematic diagram of OCPP implementation is presented in Fig.~\ref{fig:OCPP} (see Section \ref{launch_attack} for a detailed explanation). It is imperative to note that OCPP-1.6 and later versions contain functional enhancements related to smart charging that limit the maximum charging rate and on-off switching frequency \cite{alcaraz2017ocpp}. Specifically, (i) internal instructions in OCPP (\textit{TxDefaultProfile}) designates a maximum charging rate throughout the charging operation (see \underline {\textbf{CL2}} parts, Fig.~\ref{fig:OCPP}). Thus, surge attacks by increasing the charging rate (such as those proposed in \cite{sayed2022electric} and \cite{ghafouri2022coordinated}) are challenging to execute in reality. (ii) OCPP also incorporates a randomized delay in executing the instructions (\textit{RandomisedDelay} in OCPP 1.6 and \textit{SmartChargingCtrlr} in OCPP 2.0). This delay will limit the on-off switching frequency, thus limiting the effect of the coordinated switching attacks. 
While both these security measures can be bypassed if an attacker has access to the operator's internal configuration settings (i.e., defined parameters of \underline {\textbf{CL2}} parts, Fig.~\ref{fig:OCPP}), such access is usually very hard to obtain, and manipulating these settings requires CS authorization \cite{Franc2022OCPP}.

\subsection{Paper Contributions}

This work introduces a new class of attacks, named CMAs, presenting a more realistic threat to EVCS operations. To the best of our knowledge, the attack is the first to consider domain-specific features of the EVCS \emph{smart charging} environment. Executing CMAs does not require bypassing the OCPP security measures, i.e., access to commands under the privilege of a charge-point operator only and requiring authorization from the CS (i.e., the aforementioned maximum charging rate and delay functions). Furthermore, as opposed to previous studies that only consider the attack impact in terms of power grid stability, we investigate the impact of CMAs against electricity markets. Our hypothesis is that the threats against the markets are more imminent to power grid operations, as opposed to the impact on system stability, which in turn would require compromising a very large number of EVCSs\footnote{We note that inherent features in power grid operations, such as N-1 scheduling, make the system naturally resilient to load changes \cite{GoodridgeRare2023}.}. The specific contributions are as follows:


\begin{figure}[]
\centering
\includegraphics[height=6 cm,width=7 cm,trim= 10 10 490 140,clip]{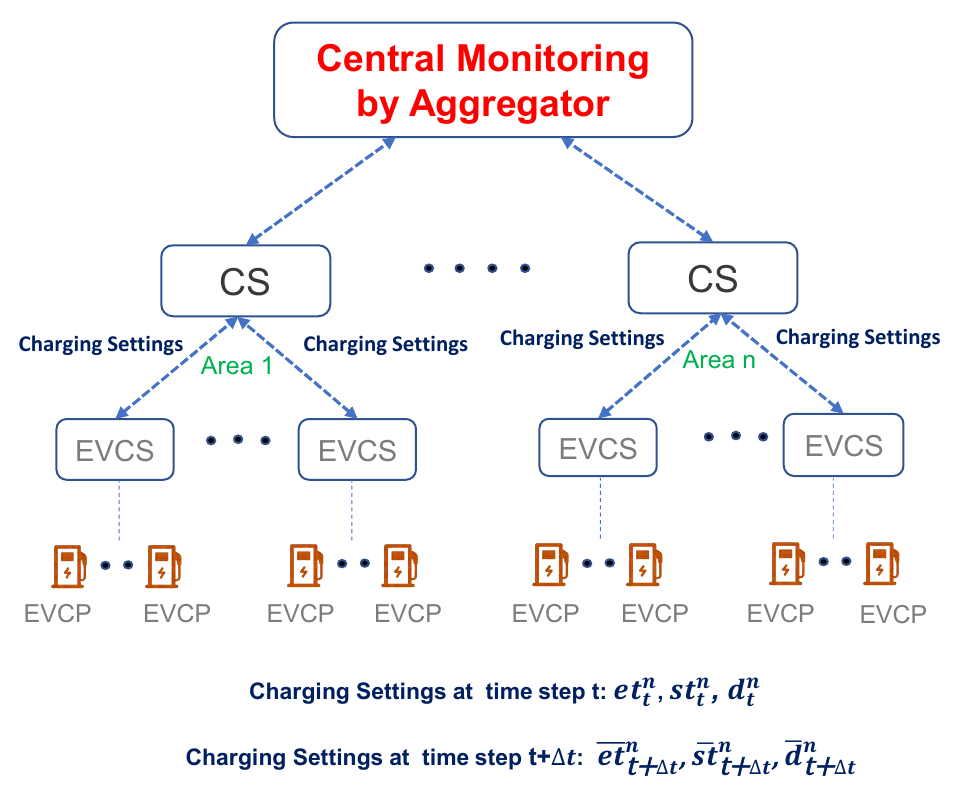}
\caption{Central Monitoring}
\label{fig:central monitoring}
\vspace{+0.2 cm}
\end{figure}

\begin{itemize}[leftmargin=*]
    \item Investigating CMAs against smart charging operations that target the energy market and demand response programs (see Section \ref{sec:attack introduction}). By leveraging the vulnerabilities in smart charging operations, the proposed CMAs can circumvent the OCPP security measures (see Section \ref{launch_attack} for more details).

    \item Considering the shifting demand, as well as the increasing demand, based on the users' preferences, and having only one-time access (in real-time) to the charging setting (unlike \cite{kabir2021two,wei2023cyber}, which requires continued access to the charging point to launch switching attacks), increases the stealthiness of these CMA surfaces; in such circumstances, it will be challenging for the CS (charging firms or aggregators), who are responsible for the smart charging task, to discern between these threats and normal fluctuations on the users' side.

    \item Presenting a new hierarchical monitoring framework (shown in Fig.~\ref{fig:central monitoring}) that uses data from the charging settings of the EVCSs (i.e., preferred start time, end time, and requested demand of the EV users) with advanced deep learning-based unsupervised anomaly detection algorithms to detect CMAs. This method is equipped with deep auto-encoders, ensuring its robustness to noisy data, a common occurrence in real-world scenarios. The integration of our modular monitoring method into existing smart charging platforms is seamless and doesn't require any additional infrastructure. 
    The monitoring framework can be viewed  as an additional layer of protection (i.e, in addition to the encryption security measures, \underline{\textbf{SL}} parts in Fig.~\ref{fig:OCPP}) that can detect various types of attacks emanating from different attack surfaces (including social engineering attacks  \cite{soykan2021disrupting}) in realtime.
    
    %

\end{itemize}

\section{Potential Cyber Threats to EVCSs and their Implications}\label{potential threats}

In this section, we commence by outlining the various sources of vulnerability in EV charging systems (i.e., EVCPs, EVCSs, and CS), supported by real-world examples. We then examine how these threats affect the aggregators (entities responsible for smart charging operations), as well as the power grid operations.

\subsection{Common Vulnerabilities in EV Charging Systems }\label{vulnerabilities}

We begin this part by enlisting a few common vulnerabilities and exposures in EV charging systems from the National Vulnerability Database (NVD)\footnote{The NVD is the U.S. government's repository of vulnerability management data based on standards and represented via the Security Content Automation Protocol (SCAP), https://nvd.nist.gov.}, given along with their NVD reference number and common vulnerability scoring system (CVSS) by January 2023.



\begin{itemize}[leftmargin=*]
    \item Server-side request forgery (CVE-2021-22821, High CVSS): allows an attacker to submit a malicious request from a susceptible server to an EV charging system, potentially gaining access to sensitive data and thereby acting on behalf of the server.  
    \item Cross-site request forgery (CVE-2022-22808, High CVSS): permits the attacker to take actions on behalf of a legitimate user without their knowledge by fooling the user's browser into submitting a request to an attacker-controlled web EV charging application. 
    \item Hard-coded credentials (CVE-2021-22730, Critical CVSS): enables the attacker to obtain unauthorized access by retrieving login credentials, such as passwords or API keys, which are directly inserted in the source code of an EV charging system for ease of coding.  
\end{itemize}

Next, we outline various vulnerabilities that attackers can exploit to breach the EV charging system and launch load-altering threats (the primary focus of this study) from EVCPs: 

\subsubsection{Software Vulnerabilities}
A seamless connection between EVCPs, EVCSs, and CSs (shown in Fig.~\ref{fig:central monitoring}) is required for reliable smart charging operating (i.e., transferring/executing real-time charging configurations).
This link is provided by various types of software, like Ampcontrol, which are vulnerable owing to their network and protocol connections \cite{ahalawat2022security}. These flaws allow an attacker to infiltrate the CSs and establish a backdoor for regaining access to launch load-altering threats (with control over a large botnet) from the charging system's core.  
\subsubsection{Hardware Vulnerabilities}
Protocols like CHAdeMo and CCS, which are widely used in high-wattage smart EVCPs (up to 400kW), are particularly susceptible to various types of cyberattacks, such as denial-of-service attacks \cite{kohler2022end}. In addition, sophisticated attackers also targeting the physical port of the EVs and EVCP (e.g., USB, etc.) These points could be accessed by unauthorized parties, such as mechanics attempting to implant malware into the charging systems.
The combination of cyber and physical attacks can have devastating effects.
The Idaho National Laboratory, for instance, did a cybersecurity analysis for DC fast charging with the CCS and CHAdeMO charging protocols. With physical and cyber access to the DC fast charging devices, they were able to control the charging of these high-wattage devices (launching a large botnet threat) \cite{hodge2019vehicle}.  

\subsubsection{Human Factors} One of the most prevalent security risks in EV charging apps is the use of weak passwords, which can enable attackers to launch attacks from the users' end. The effectiveness of the security update patches for EV charging apps is largely driven by the users' behavior, as some part of the users and insiders fail to update their systems in time. Furthermore, social engineering attacks on EVCSs, can involve influencing individuals to provide sensitive information or undertake acts that may jeopardize the EVCSs or the EV charging network's security \cite{soykan2021disrupting}.
Phishing, baiting, pretexting, and tailgating are examples of such tactics. Although providing regular security training and awareness programs for EV charging companies' customers and personnel can help to mitigate the consequences of these threats, there is always a security risk on the end-user side that can empower adversaries to launch load-altering threats. 



\subsection{Implications of Attacks on Aggregators and ISOs}

Load forecasting is a crucial aspect of managing modern power grids, as it provides valuable insights into the expected levels of energy consumption at different times and locations. The accuracy of load forecasting results will be significantly impacted by load-altering threats -- including such shifting and increasing/decreasing peak demand, the primary impact of the stealthy CMAs from EVCSs outlined in this work. In this section, we discuss the impact of the aforementioned threats on modern power grids from various perspectives.

\subsubsection{Power Grid Operations} 
Demand response programs aim to reduce electricity usage during periods of high demand by encouraging consumers to shift their consumption to off-peak hours. With the rise of the Internet of Things (IoT), these programs are expected to become more common in energy management activities in smart grids. In 2018, U.S. utilities used demand response services to reduce/shift approximately 4.5\% of their peak load capacity; it is projected that this figure will increase to 20\% by 2030, resulting in annual cost savings of over \$15 billion \cite{acharya2021causative}. In addition, demand response programs can facilitate the integration of renewable energy resources by enhancing grid stability and reliability, which is crucial for achieving Net-Zero aims. Large-scale CMAs with EVCSs can disrupt demand response programs by creating unexpected demand in the load. Such attacks can also become an unforeseen contingency in unit commitment programs where the operator plans the hourly generation schedule to supply the forecasted load over a look-ahead horizon \cite{shayan2019network}. 

\subsubsection{Energy Market}\label{energy_market}
Accurate load forecasting in the energy market can yield significant profits for the market participants, including aggregators who are tasked with managing smart charging for EVs, shown in Fig.~\ref{fig:central monitoring}. In contemporary energy markets such as CAISO, there are two primary bidding phases \cite{jahangir2021novel}: (i) Day-Ahead (DA) Bidding: Market participants, including aggregators (such as firms responsible for smart charging of EVs), submit bids and offers for electricity in the day-ahead market (DAM), typically one day prior to the delivery period. These bids indicate the quantity and price of electricity they intend to purchase based on the anticipated demand conditions for the following day. (ii) Real-Time (RT) Bidding: Once the DAM clears\footnote{An auction-based process where bids for buying/selling energy in DAM are cleared to determine accepted offers.} and establishes locational marginal prices (LMPs), market participants, including aggregators, modify their bids and offers in the real-time market (RTM) based on actual market conditions, such as load demand changes and other real-time factors. The aggregator may face the consequences for failing to fulfill their bids in RTM in the form of financial penalties and loss of market access. Financial penalties are usually calculated based on the difference between the price of the unfulfilled bid and the RTM price during the period when the bid was not met. Additionally, if unfulfilled bids continue to accumulate, the aggregator may lose access to the market for a certain period of time, preventing them from trading in future DAM or RTM until the issue is resolved. Consequently, large-scale botnet attacks from EVCSs (such as Manipulation of Demand via IoT, called MaDIoT attacks \cite{shekari2021mamiot}) that are able to manipulate the demand profiles can cause severe problems for aggregators in the energy markets.

\section{Stealthy CMAs against EVCSs}\label{sec:attack introduction}


\begin{figure}[]
\centering
\includegraphics[height=5.5 cm,width=8.5 cm,trim= 110 10 110 68,clip]{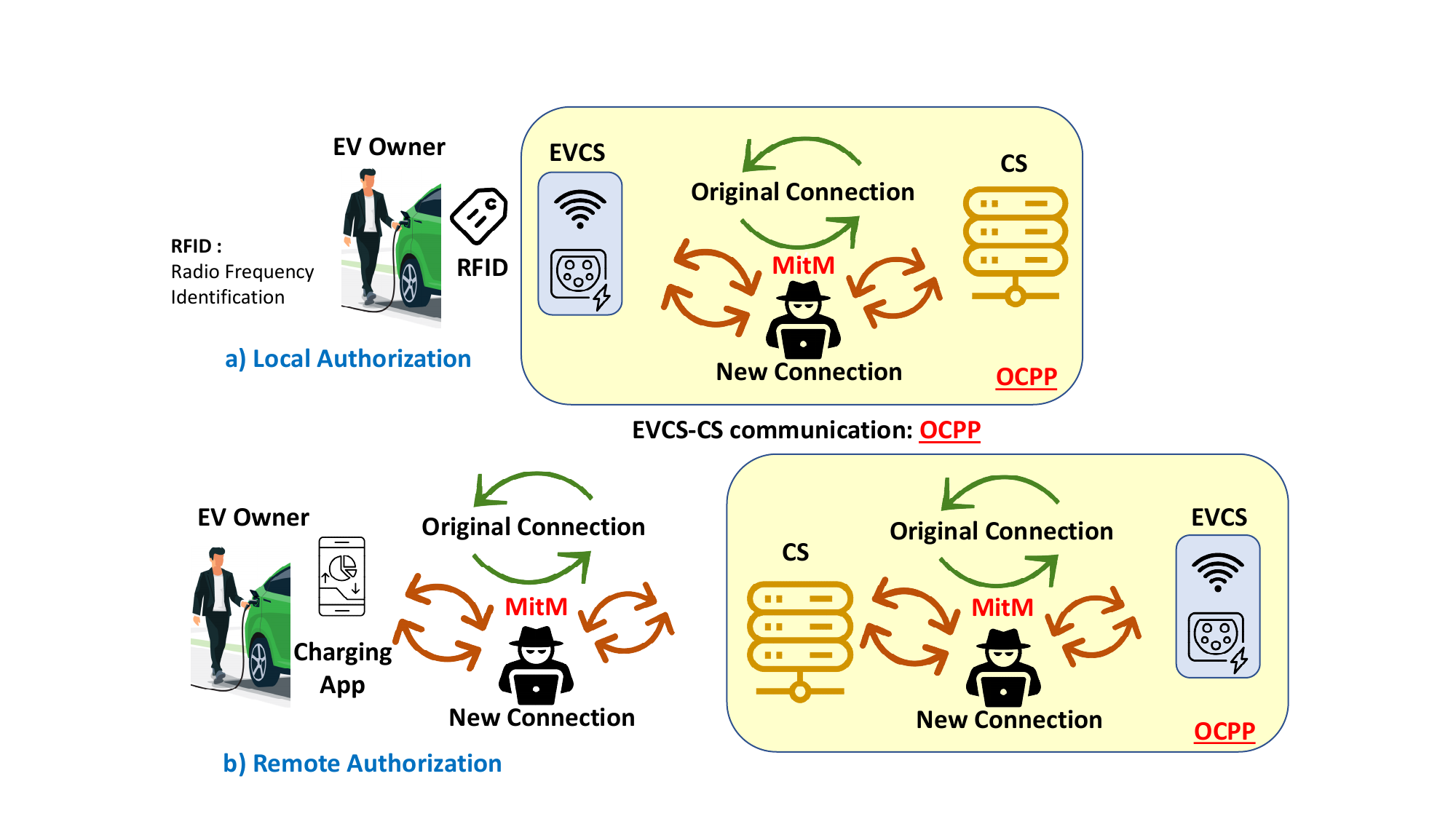}
\caption{MitM attacks on smart charging.}
\label{fig:MitM}
\vspace{+0.2 cm}
\end{figure}

In this section, we delve into the general process of designing stealthy CMAs within EVCSs. We begin by elucidating the two potential authentication scenarios in smart charging tasks, as well as providing relevant details on selecting the target EVCSs and obtaining control over them. Subsequently, we expound on the procedure for launching different types of CMAs in the charging environments.

\subsection{Locate the Target EVCSs and Seize their Control}

As depicted in  Fig.~\ref{fig:MitM}, in a smart charging environment, when an EV user plugs their vehicle into the EVCS  they have the option to specify their charging preferences, including the preferred start time, planned departure time (end time), and requested demand. These preferences can be set manually via EVCPs (local authentication in Fig.~\ref{fig:MitM}) or through a smartphone application (remote authentication in Fig.~\ref{fig:MitM}). CS then aggregates the charging preferences of multiple users connected to EVCSs in a particular geographic area. The CS executes a smart charging scheme to schedule the charging operations, determining the charging rate of each EVCP during a specific time period, based on a variety of factors, including technical constraints, electricity price cost, and planned demand constraints in the energy market (which were briefly discussed in Section~\ref{energy_market} and will be explained in the detailed formulation in Section~\ref{market:formulation}). 

As described in Section~\ref{vulnerabilities}, the vulnerabilities in the communication between EVs, EVCSs, and CS can be exploited by attackers. By gaining access to the charging settings, attackers can manipulate the requested charging demands to achieve their objectives using MitM attacks, as illustrated in Fig.~\ref{fig:MitM}. Further details on this attack method will be discussed in the subsequent section. In order to carry out efficient attacks, attackers must be able to acquire both topological data (revealing the connection of EVCSs to specific power grid nodes, as well as the power profile of these nodes) and technical data (such as the total number of EVCPs at each EVCS and their nominal power rates) pertaining to the targeted EVCSs; reference \cite{acharya2020public} offers an elaborate analysis of this matter.

\subsection{Launch CMAs}\label{launch_attack}

\begin{figure}[]
\centering
\includegraphics[height=7 cm,width=8.5 cm,trim= 110 60 140 0,clip]{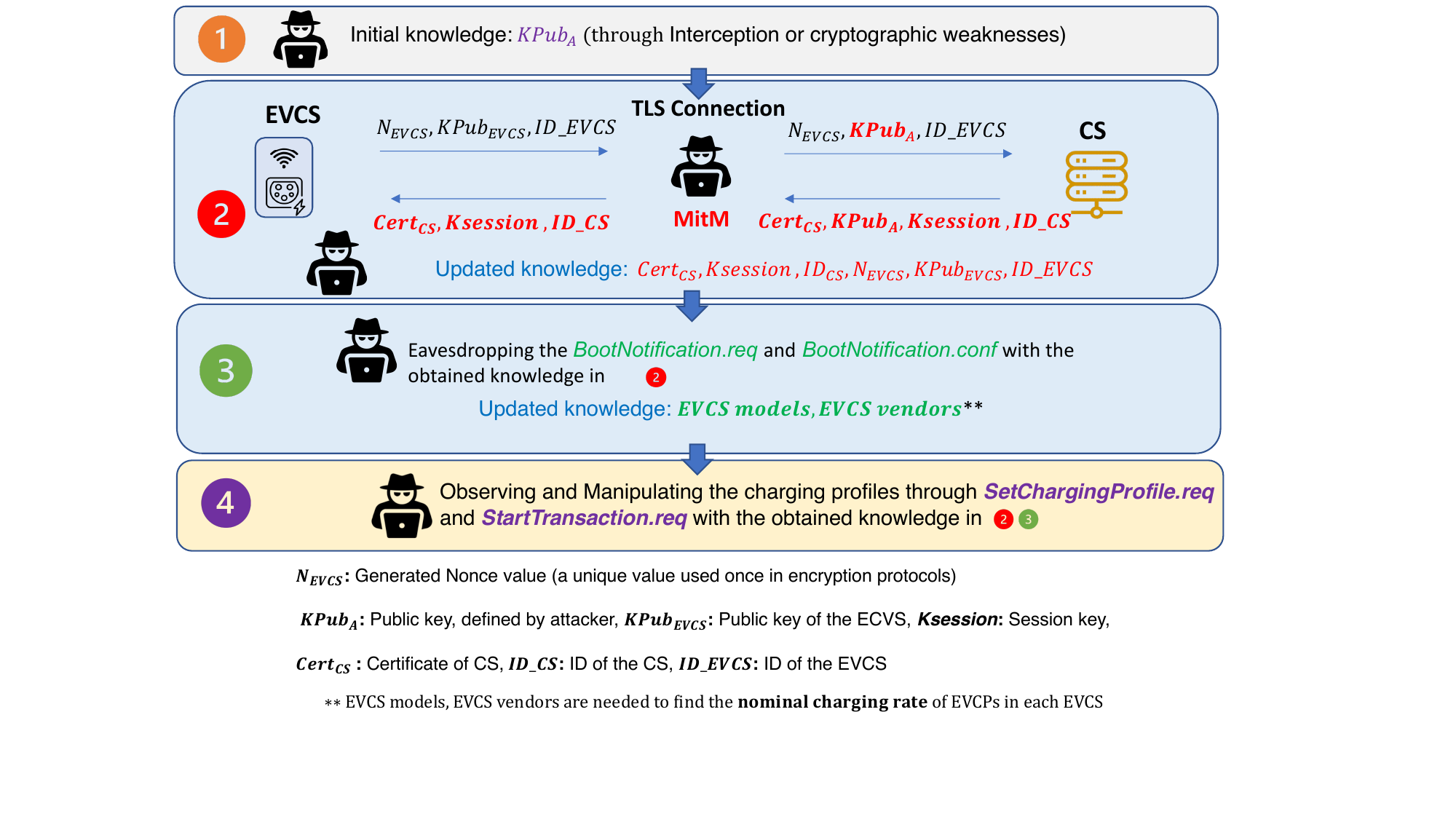}
\caption{
The overall process of launching CMAs on EVCSs}
\label{fig:launching}
\vspace{+0.2 cm}
\end{figure}

\begin{figure}[]
\centering
\includegraphics[height=7 cm,width=8.8 cm,trim= 5 80 190 80,clip]{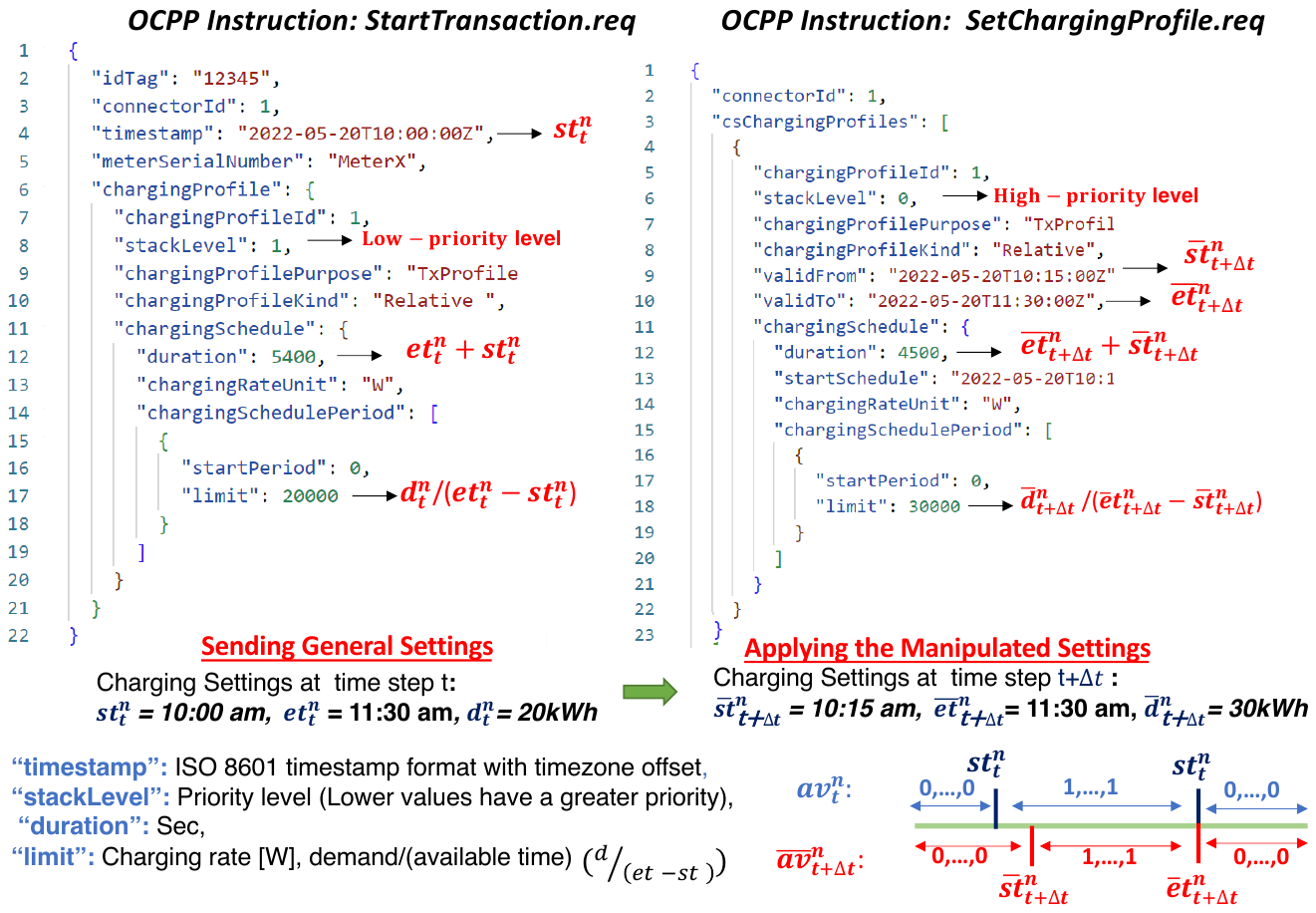}
\caption{
Example JSON-formatted OCPP instructions for sending general charging settings (in this example, $st_{t}^{n}=10$ a.m., an $et_{t}^{n}=11:30$ a.m., and $d_{t}^{n}=20 kWh$) using the \textit{StartTransaction.req} instruction and applying manipulated charging settings ($st_{t+\Delta t}^{n}=10:15$ a.m., an $et_{t+\Delta t}^{n}=11:30$ a.m., and $d_{t+\Delta t}^{n}=30 kWh$; this is a sample of \textit{Type-4 CMAs}, presented in \textbf{Algorithm \ref{algorithm}}) with the \textit{SetChargingProfile.req} instruction.}

\label{fig:codes}
\vspace{+0.2 cm}
\end{figure}

\textbf{Algorithm \ref{algorithm}} outlines the key steps of the proposed CMAs on EVCSs, providing a high-level overview. The algorithm takes the charging parameters of the targeted EVCPs (i.e., $\{ st_{t}^{n} \}_{t \in \mathcal\mathcal{T_a}, {n \in \mathcal\mathcal{N}}}$, $\{ et_{t}^{n} \}_{t \in \mathcal\mathcal{T_a}, {n \in \mathcal\mathcal{N}}}$, $\{ d_{t}^{n} \}_{t \in \mathcal\mathcal{T_a}, {n \in \mathcal\mathcal{N}}}$, defined by the EV owners) and attack control parameters (i.e.,  $C_{att}$, $Ch_{av}$,  $\left | \mathcal{N} \right |$, $\left | \mathcal{N}_h \right |$, $\left | \mathcal{N}_{hd} \right |$, $\left | \mathcal{N}_{hs} \right |$, $\left | \mathcal{N}_{he} \right |$, $Pl$, $\Delta t$, defined by the attackers) as input data and generates manipulated charging settings (i.e., $\{\overline{st}_{t+\Delta t}^{n}\}_{t+\Delta t \in \mathcal\mathcal{T_a}, {n \in \mathcal\mathcal{N}}}$, $\{\overline{et}_{t+\Delta t}^{n}\}_{t+\Delta t \in \mathcal\mathcal{T_a}, {n \in \mathcal\mathcal{N}}}$,  $\{\overline{d}_{t+\Delta t}^{n}\}_{t+\Delta t \in \mathcal\mathcal{T_a}, {n \in \mathcal\mathcal{N}}}$) as the output results. 
Accessing the charging setting profiles (i.e., input data of \textbf{Algorithm \ref{algorithm}}) and injecting manipulated charging setting profiles (i.e., the output of \textbf{Algorithm \ref{algorithm}}) into EVCS (within Local authentication, Fig.~\ref{fig:MitM}) or CS\footnote{As illustrated in Fig.~\ref{fig:MitM}, manipulated charging profiles can also be injected into EVCSs via EV charging applications in the Remote authentication task.} (within Remote authentication, Fig.~\ref{fig:MitM}) can be accomplished during the execution of \textit{StartTransaction.req} and \textit{SetChargingProfile.req}  instructions in the OCPP environment (\underline {\textbf{CL1}} in Fig.~\ref{fig:OCPP}). To carry out these CMAs, as shown in Fig.~\ref{fig:launching}, the attackers must intercept the communication channels (details mentioned in Section~\ref{vulnerabilities}). By exploiting vulnerabilities in the TLS mechanisms (through \textit{CommTLS.init} and \textit{CommTLS.resp} instructions) and boot Notifications (through \textit{BootNotification.req} and \textit{BootNotification.conf} instructions) in OCPP. This involves gaining access to the credentials of an OCPP user (\textit{public}, \textit{private}, and \textit{session} keys), \underline {\textbf{SL}}, Fig.~\ref{fig:OCPP}.
A demonstration of data manipulation, outlined in Part 4 of Fig.~ \ref{fig:launching}, is depicted in Fig~\ref{fig:codes}.

As depicted in Fig.~\ref{fig:launching} and Fig.~\ref{fig:codes}, unlike the \textit{surge in demand} and \textit{coordinated switching} attacks, discussed in Section \ref{survey}, these threats do not require circumventing the additional controlling settings in OCPP, particularly those related to the maximum charging rate (\textit{TxDefaultProfile}) and random delays (\textit{RandomisedDelay} or \textit{SmartChargingCtrlr}) in performing charging profiles, and can be conducted in a more covert manner compared to them. Further details on the feasibility and techniques for executing these attempts can be found in \cite{alcaraz2017ocpp}. Following is a detailed explanation of the stealthy attacks introduced in \textbf{Algorithm \ref{algorithm}}:\\


\begin{algorithm}[t]
\footnotesize
	\caption{ Launching CMAs on smart EVCPs}
    \underline{Input:} Charging profiles: $\{ st_{t}^{n} \}_{t \in \mathcal\mathcal{T}_a, {n \in \mathcal\mathcal{N}}}$, $\{ et_{t}^{n} \}_{t \in \mathcal\mathcal{T}_a, {n \in \mathcal\mathcal{N}}}$, $\{ d_{t}^{n} \}_{t \in \mathcal\mathcal{T_a}, {n \in \mathcal\mathcal{N}}}$, Controlling parameters: $C_{att}$, $Ch_{av}$, $Pl$, $\Delta t$ , $\left | \mathcal{N} \right |$, $\left | \mathcal{N}_h \right |$, $\left | \mathcal{N}_{hd} \right |$, $\left | \mathcal{N}_{hs} \right |$, $\left | \mathcal{N}_{he} \right |$, $\mathcal{N}_{h} = ({\mathcal{N}_{hd} \cup \mathcal{N}_{hs} \cup \mathcal{N}_{he}}) \subset \mathcal{N}$  \\
    \underline{Output:} Charging profiles: $\{\overline{st}_{t+\Delta t}^{n}\}_{t+\Delta t \in \mathcal\mathcal{T_a}, {n \in \mathcal\mathcal{N}}}$, $\{\overline{et}_{t+\Delta t}^{n}\}_{t+\Delta t \in \mathcal\mathcal{T_a}, {n \in \mathcal\mathcal{N}}}$,  $\{\overline{d}_{t+\Delta t}^{n}\}_{t+\Delta t \in \mathcal\mathcal{T_a}, {n \in \mathcal\mathcal{N}}}$ \\
    \textbf{Defining initial parameters for calculating} $Av_t^{n}, Ch_t^{n}$:
    
	\begin{algorithmic}[1]
	\STATE $n_{step} \gets Pl(60/\Delta t)$\\
	\STATE $Av_t^{n} \gets Zeros\left [  1,n_{step} \right ]$\\
	\FOR{ $\left | \mathcal{N} \right |$ iteration}
        \IF {$st_{t}^{n} < et_{t}^{n}$, (charging in a day)}
        \STATE {$Av_t^{n} \left [st_{t}^{n}, et_{t}^{n} \right ] \gets 1$}
	\ELSE 
        \STATE{$Av_t^{n} \left[ 1, et_{t}^{n} \right] \gets 1,
                   Av_t^{n} \left [st_{t}^{n}, n_{step} \right ] \gets 1$ (more than a day)}
	\ENDIF\\
        \STATE {$Ta^n_t \gets \sum_{t=1}^{\left | \mathcal{T} \right |}Av_t^{n}$}
        \STATE {$Ch_t^{n} \gets  d_{t}^{n}(60/\Delta t)/Ta^n_t $}
	\ENDFOR\\
       \textbf{Launching Type-1 CMAs:}\\
         \STATE $ACR_{att} \gets C_{att} \times Ch_{av}$\\
	\FOR{ $\left | \mathcal{N}_{hd} \right |$ iteration (\textbf{Attack type 1})}
	\STATE Randomly select a EVCP (${n}_{hd} \in \mathcal{N}_{hd}$ $ \subset \mathcal{N}_h $)\\
        \STATE $\overline{st}_{t+\Delta t}^{{n}_{hd}} \gets {st}_{t}^{{n}_{hd}}$\\
        \STATE $\overline{et}_{t+\Delta t}^{{n}_{hd}} \gets {et}_{t}^{{n}_{hd}}$\\
        \STATE $\overline{Av}_{t+\Delta t}^{{n}_{hd}} \gets {Av}_{t}^{{n}_{hd}}$\\
        \STATE $\overline{Ta}_{t+\Delta t}^{{n}_{hd}} \gets {Ta}_{t}^{{n}_{hd}} $\\
	  \STATE $\overline{d}_{t+\Delta t}^{{n}_{hd}} \gets ACR_{att}~Ta_t^{{n}_{hd}} (\Delta t/60)- Ch_t^{{n}_{hd}} 
               (\Delta t/60) + {d}_{t}^{{n}_{hd}}$\\
	\STATE $\overline{Ch}_{t+\Delta t}^{{n}_{hd}} \gets  \overline{d}_{t+\Delta t}^{{n}_{hd}}(60/\Delta        t)/Ta^{{n}_{hd}}_t $ \\
	\ENDFOR
	    
	\FOR{ $\left | \mathcal{N}_{hs} \right |$ iteration (\textbf{Launching Type-2 CMAs})}
	\STATE Randomly select a EVCP (${n}_{hs} \in \mathcal{N}_{hs}$ $ \subset \mathcal{N}_h $)
	\STATE      $\overline{d}_{t+\Delta t}^{{n}_{hs}} \gets ACR_{att}~Ta_t^{{n}_{hs}} (\Delta t/60) - Ch_t^{{n}_{hs}} (\Delta t/60) + {d}_{t}^{{n}_{hs}}$
  	\STATE    $\overline{Ch}_{t+\Delta t}^{{n}_{hs}} \gets  \overline{d}_{t+\Delta t}^{{n}_{hs}}(60/\Delta t)/Ta^{{n}_{hs}}_t $ 
  	\STATE    $\overline{Ta}_{t+\Delta t}^{{n}_{hs}} \gets  {d}_{t}^{{n}_{hs}}(60/\Delta t)/\overline{Ch}_{t+\Delta t}^{{n}_{hs}} $
	             
        \STATE  $\overline{st}_{t+\Delta t}^{{n}_{hs}} \gets {st}_{t}^{{n}_{hs}} + \left | {Ta}_{t}^{{n}_{hs}} - \overline{Ta}_{t+\Delta t}^{{n}_{hs}}  \right |$
	\STATE  $\overline{et}_{t+\Delta t}^{{n}_{hs}} \gets {et}_{t}^{{n}_{hs}}$
	\STATE $\overline{Av}_{t+\Delta t}^{{n}_{hs}} \gets$ updating by lines (4-8) with $\overline{st}_{t+\Delta t}^{{n}_{hs}}, \overline{et}_{t+\Delta t}^{{n}_{hs}} $
         \ENDFOR

    \FOR{ $\left | \mathcal{N}_{he} \right |$ iteration (\textbf{Launching Type-3 CMAs})}
       \STATE Randomly select a EVCP (${n}_{he} \in \mathcal{N}_{he} \subset \mathcal{N}_h $ )
              \STATE$\overline{d}_{t+\Delta t}^{{n}_{he}} \gets ACR_{att}~Ta_t^{{n}_{he}} (\Delta t/60) - Ch_t^{{n}_{he}} (\Delta t/60) + {d}_{t}^{{n}_{he}}$
            \STATE$\overline{Ch}_{t+\Delta t}^{{n}_{he}} \gets  \overline{d}_{t+\Delta t}^{{n}_{he}}(60/\Delta t)/Ta^{{n}_{he}}_t $ 
            \STATE$\overline{Ta}_{t+\Delta t}^{{n}_{he}} \gets  {d}_{t}^{{n}_{he}}(60/\Delta t)/\overline{Ch}_{t+\Delta t}^{{n}_{he}} $
              \STATE$\overline{et}_{t+\Delta t}^{{n}_{he}} \gets {et}_{t}^{{n}_{he}} - \left | {Ta}_{t}^{{n}_{he}} - \overline{Ta}_{t+\Delta 
                t}^{{n}_{he}}  \right |$
              \STATE$\overline{st}_{t+\Delta t}^{{n}_{he}} \gets {st}_{t}^{{n}_{he}} $
              
              \STATE$\overline{Av}_{t+\Delta t}^{{n}_{he}} \gets$ updating by lines (4-8) with $\overline{st}_{t+\Delta t}^{{n}_{he}}, \overline{et}_{t+\Delta t}^{{n}_{he}} $\\
    \ENDFOR
	     \STATE Combination of lines (13-21) and (22-30), $\left | \mathcal{N}_{hd} \right |, \left | \mathcal{N}_{hs} \right | \geq 0.4 \times \left | \mathcal{N}_{h} \right |$ (\textbf{Launching Type-4 CMAs})
	     \STATE Combination of lines (13-21) and (31-39), $\left | \mathcal{N}_{hd} \right |, \left | \mathcal{N}_{he} \right | \geq 0.4 \times \left | \mathcal{N}_{h} \right |$ (\textbf{Launching Type-5 CMAs})
	     \STATE Mix of the combination of lines (13-21), (22-30) and (31-39) $\left | \mathcal{N}_{hd} \right |, \left | \mathcal{N}_{he} \right |, \left | \mathcal{N}_{hs} \right | \leq 0.4 \times \left | \mathcal{N}_{h} \right |$ (\textbf{Launching Type-6 CMAs})
	
\end{algorithmic}
\label{algorithm}
\end{algorithm}

\textbf{\textit{Initializing coordinated charging parameters (lines 1-11): }} 
Smart charging involves three critical parameters that are submitted by an EV user once their vehicle is plugged into the EVCP \cite{lee2019acn} -- (i) a start time ($st_{t}^{n}$), (ii) end time ($et_{t}^{n}$), and (iii) the requested demand ($d_{t}^{n}$ [kWh]). The smart charging process sets the charging rate for EVCP $n$ ($Ch_t^{n}$ [kW]) by dividing the requested demand by the total duration of time for which the EV is connected to the EVCP. Lines 1-11 of \textbf{Algorithm \ref{algorithm}} specify the coordinate charging process \footnote{For ease of illustration, here, we examine a simplified version of the coordinated charging, while Section~\ref{market:formulation} presents an advanced smart charging approach in the energy market environment to validate the impacts of these attacks.}, which we explain in the following.
First, we assume a slotted time system with a slot duration specified by $\Delta t$ (expressed in minutes) as shown in Fig.~\ref{fig:codes} (timeline shown at the bottom) and a planning horizon $Pl$ (usually a 24-hour period).
The number of charging steps ($n_{step}$) is then calculated as in line 1.  We then create a vector $Av_t^{n}$ of ones and zeros of length $(60/\Delta t) \times Pl$, with ones representing the time slots (within the sampling duration) for which an EVCP is occupied and zeros denoting non-occupancy. Note that $(60/\Delta t) \times Pl$ represents the number of time slots during the planning horizon. 
Assume that an EV is plugged in for charging with a start time of ($st_{t}^{n}$) and end time 
 of ($et_{t}^{n}$). Accordingly, we set the occupancy slots within the vector $Av_t^{n}$ to $1$ depending on whether the charging task is within a single day (line 5) or takes more than a day (line 7). The total charging time in terms of the time slots ($Ta^n_t$ [$\Delta t$-min]) is calculated as the sum of all elements in the $Av_t^{n}.$ Line 10 sets the charging rate for EVCP $n.$ 



\textbf{\textit{Setting the Attack Parameters (line 12 and Input):}} We consider false data injection attacks that target the charging parameters (start/end time and requested demand). We assume that the attacker observes the true values of these parameters at time $t,$ and injects the manipulated parameters during the next time slot $t + \Delta t.$

At the outset, the attacker determines the magnitude of increase in the charging rate ($ACR_{att} $) that they wish to subject the EVCSs to depending on their objective. 
In order to preserve the stealthy aspect of the attacks, we set this quantity to be less than 10\% of the average charging rate across all of the targeted EVCPs ($C_{att}\leq0.1$). 

   \textbf{\textit{Launching Type-1 CMAs (lines 13-21):}} First, $\left | \mathcal{N}_{hd} \right |$ number of compromised EVCPs ($\mathcal{N}_{hd}$ $ \subset \mathcal{N}_h $) are chosen by the adversary and their requested demands of the users connected to them ($d_{t}^{n}$) are manipulated, while other charging settings stay the same as before (${st}_{t}^{{n}_{hd}}$, ${et}_{t}^{{n}_{hd}}$). The manipulated demand value ($\overline{d}_{t+\Delta t}^{{n}_{hd}}$ [kWh]) is then calculated in line~19 based on the increased charging rate under attack $ACR_{att}$ [kW] (first term in the right-hand side) while subtracting the charge that has already occurred between $t$ and $t+\Delta t$ (second-term in the right-hand side) and added to the originally requested demand. Finally, the new charging rate ($\overline{Ch}_{t+\Delta t}^{{n}_{hd}}$) 
   is calculated in line~20. 

   




\textbf{\textit{Launching Type-2 CMAs (lines 22-30):}} The primary goal, in this case, is to readjust the EVCPs' start times in order to shift the peak demand to later hours (i.e., shift right) while raising the requested demand. First, $\left | \mathcal{N}_{hs} \right |$ number of compromised EVCPs ($ \mathcal{N}_{hs}$ $ \subset \mathcal{N}_h $) are chosen by the adversary to manipulate the requested start time of users (${st}_{t}^{{n}_{hs}}$), while end time (${et}_{t}^{{n}_{hs}}$) stays the same as before. Similar to Type-1 CMAs, the manipulated demand  ($\overline{d}_{t+\Delta t}^{{n}_{hs}}$ [kWh]) and the new charging rate ($\overline{Ch}_{t+\Delta t}^{{n}_{hs}}$) is calculated in lines~24 and 25 respectively. 
As the attacker's aim is to shift the start time for the selected EVCPs (${st}_{t}^{{n}_{hs}}$), the manipulated 
available time ($\overline{Ta}_{t+\Delta t}^{{n}_{hs}}$) is calculated as in line~26. Note that this expression forces the smart charging system to satisfy the original requested time with a higher charging rate (i.e., the manipulated value), thus reducing the available time. Following this, the new start time ($\overline{st}_{t+\Delta t}^{{n}_{hs}}$) is calculated as in line~27,  effectively shifting the charging operation to later periods. Lastly, the new total available time vector ($\overline{Av}_{t+\Delta t}^{{n}_{hs}}$) is computed as in line~29.




\textbf{\textit{Launching Type-3 CMAs (lines 31-39):}} In this scenario, the major objective is to modify the EVCPs' end time in order to shift peak demand to earlier hours (i.e., shift left) while simultaneously increasing the requested demand. The steps followed in this attack are enlisted in lines 31-39, which are similar to Type-2 CMAs, except that we manipulate the end time (${et}_{t}^{{n}_{he}}$) in line 36 while keeping start time (${st}_{t}^{{n}_{he}}$) the same.   

\textbf{\textit{Mixed CMAs, Types (4-6) (lines 40-42):}} The previous CMAs assume that the attacker subjects all the EVCPs that are under their control to the same type of attack. Under mixed CMAs, we provide attackers with the flexibility to launch a combination of these CMAs. Specifically, of the $\mathcal{N}_h$ EVCPs under the control of the attacker, we assume that the attacker injects Type-1 CMAs into $\mathcal{N}_{hd} (\subseteq \mathcal{N}_h)$ EVCPs, Type-2 CMAs into $\mathcal{N}_{hs} (\subseteq \mathcal{N}_h)$ and Type-3 CMAs into $\mathcal{N}_{he} (\subseteq \mathcal{N}_h).$  We further divide the mixed attacks into three categories as in lines 40-43, where the CMA types differ in terms of the size of the subsets $\mathcal{N}_{hd}, \mathcal{N}_{hs} \ \text{and} \ \mathcal{N}_{he}.$



\begin{figure}[!h]
\centering
\includegraphics[height=5 cm,width=8.8 cm,trim= 100 5 100 20,clip]{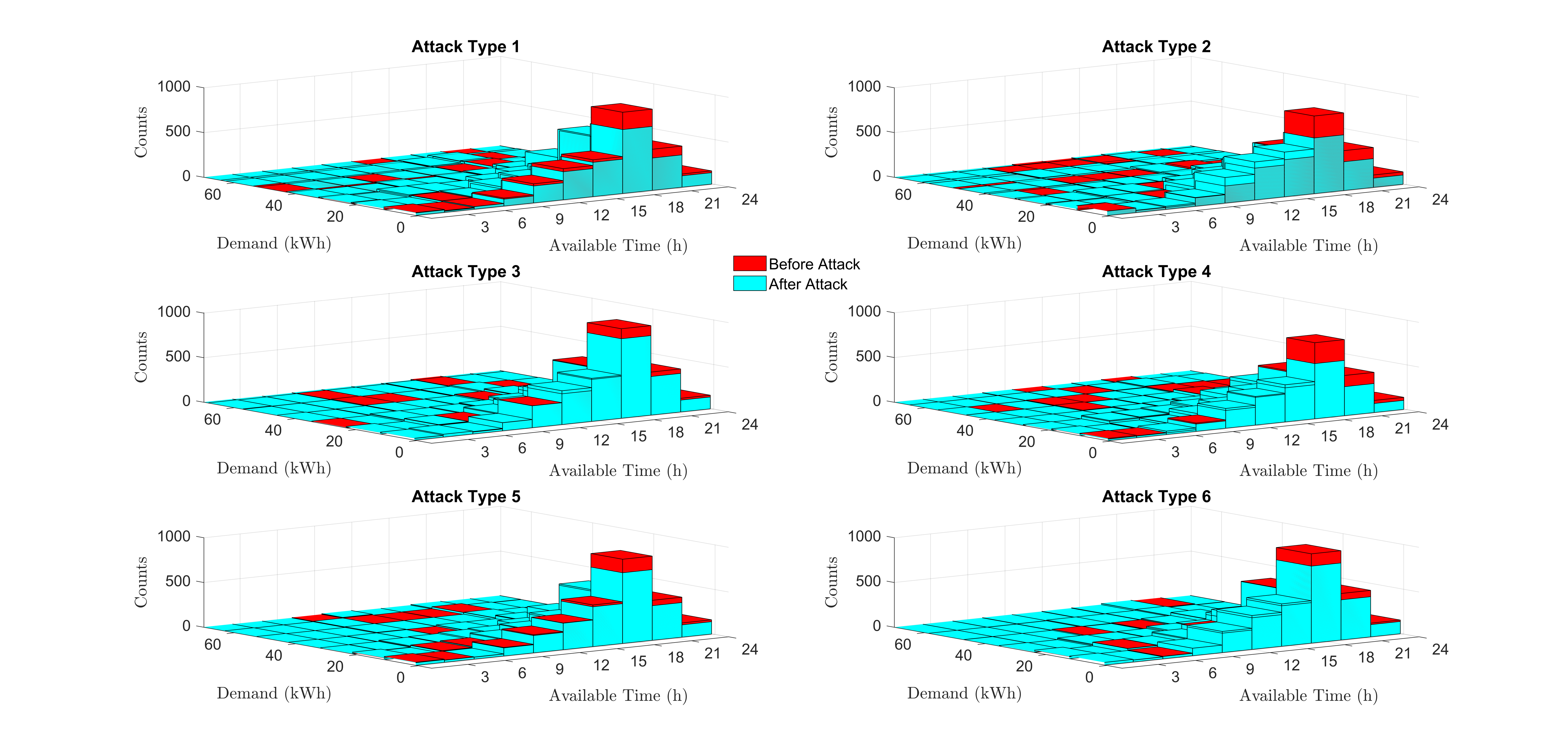}
\caption{3D histogram of all CMA types, 5000 EVCPs}
\label{fig:3D hist_all attack types}
\vspace{-0.2 cm}
\end{figure}

\begin{figure}[!h]
\centering
\includegraphics[height=5 cm,width=8.8 cm,trim= 80 5 100 5,clip]{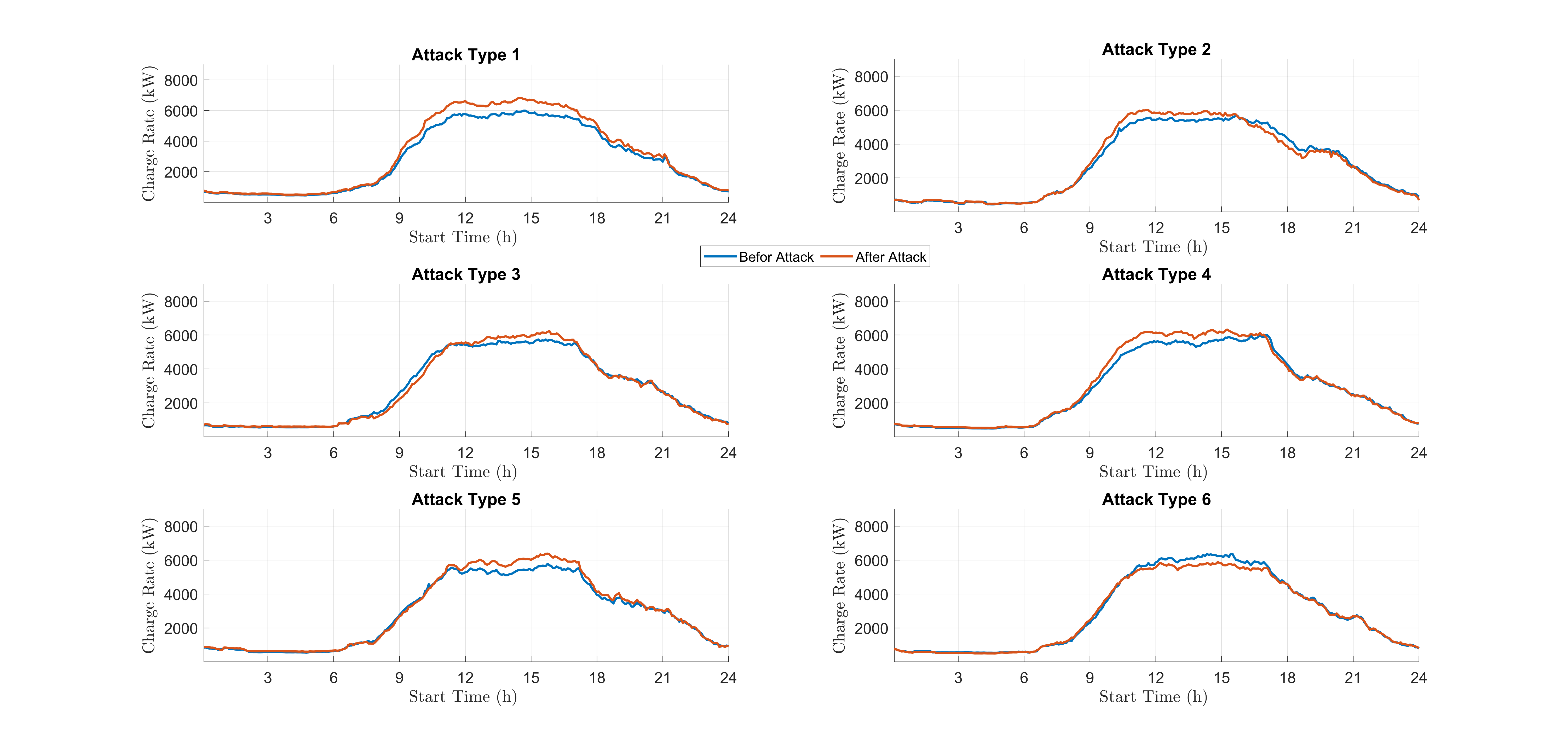}
\caption{Charging profile of all CMA types, 5000 EVCPs}
\label{fig:Charge_all attack types}
\vspace{-0.2 cm}
\end{figure}

Fig. \ref{fig:3D hist_all attack types} and Fig. \ref{fig:Charge_all attack types} provide a visual representation of the discussed CMAs, displaying 3D histograms of the charging settings and coordinated charging profiles for both before and after the initiation of the CMA scenarios -- attack control parameters: $C_{att}= 0.08$, $Ch_{av} = 30kW$,  $\left | \mathcal{N} \right | = 5000 $, $\left | \mathcal{N}_h \right | = 0.7 \times \left | \mathcal{N} \right |$, $Pl = 24 h$, $\Delta t = 5~ min$.

\section{Detection of CMAs on EVCSs with  deep learning-based monitoring approaches}\label{sec:detection scheme}

\begin{figure}[]
\centering
\includegraphics[height=6 cm,width=9 cm,trim= 2 60 15 70,clip]{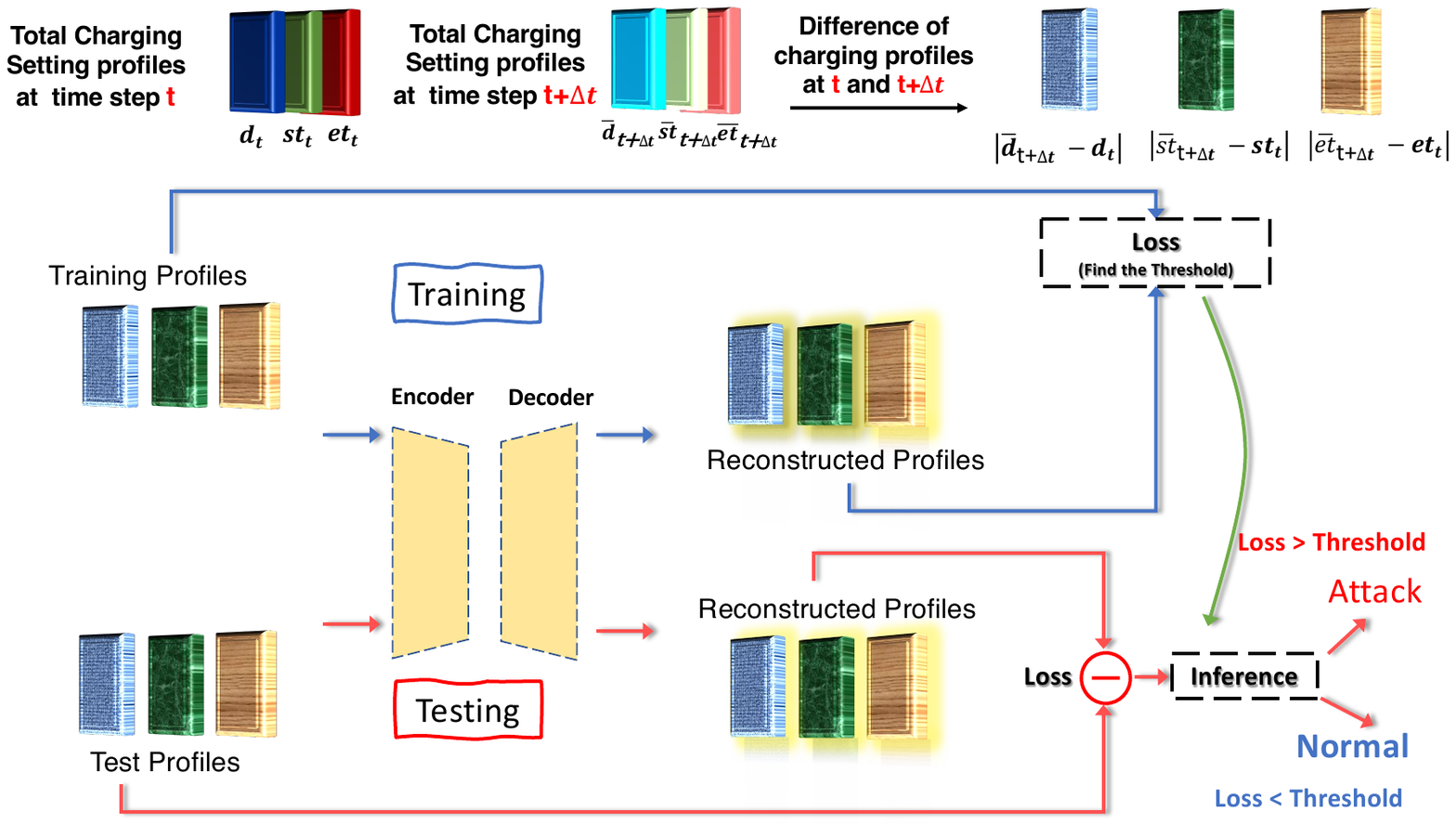}
\caption{
Proposed data-driven monitoring framework using deep auto-encoder for detecting CMAs}
\label{fig:AE_detection}
\vspace{+0.2 cm}
\end{figure}

In this section, we present a data-driven monitoring framework designed to safeguard EVCSs against potential CMAs, described in \textbf{Algorithm \ref{algorithm}}. To achieve this goal, we introduce a modular monitoring framework, shown in Fig.~\ref{fig:AE_detection}, that seamlessly integrates with cloud-based smart charging platforms (depicted in Fig.~\ref{fig:central monitoring}), eliminating the need for additional measurements or communication infrastructure. It is important to emphasize that our proposed security solution primarily focuses on the charging profiles of EVCSs, such as start time, end time, and requested demand; additional profiles can be incorporated based on different use cases. The proposed security solution can be seen as an extra layer of protection that can be implemented alongside existing encryption-based security solutions.
Our data-driven monitoring framework utilizes an unsupervised anomaly detection auto-encoder architecture based on a 2D-CNN structure (Table \ref{tab:CNN-AE}).\footnote{
2D-CNNs are generally faster than 1D-CNNs when dealing with large datasets, and they possess the ability to identify patterns irrespective of their location in monitoring data. On the other hand, 1D CNNs are more reliant on the ordering of the input data and can be sensitive to it.} The unsupervised learning approach overcomes the limitations of limited labeled data, allowing for anomaly detection without relying on pre-labeled examples. This adaptability is particularly valuable in scenarios where anomaly patterns are unknown or constantly evolving. 

As depicted in Fig.~\ref{fig:AE_detection}, the input data for the anomaly detection framework is obtained by computing the difference between the charging profiles at time step $t$ and $t+\Delta t$ using a structure of $\left | \mathcal{N} \right | \times 3$ (where $\left | \mathcal{N} \right |$ represents the number of monitored EVCPs and $3$ is the number of data sources that must be monitored, i.e., $st, et, d$). In order to utilize a 2D-CNN structure, as given in Table \ref{tab:CNN-AE}, the data is reshaped into a structure of $N_x \times 100 \times 3$, where, $N_x = \left | \mathcal{N} \right |/100$, before applying the auto-encoder. By training the autoencoder on a dataset of normal samples, it learns to reconstruct them accurately. Anomalies can then be detected by measuring the difference between the original sample and its reconstruction, using the Binary cross-entropy loss function (Equation.~\ref{eqn:BCE loss}). The reconstruction error is a measure of dissimilarity, where a higher error indicates a higher likelihood of an anomaly. To make decisions on whether a monitoring sample contains an anomaly, a threshold needs to be defined (Equation.~\ref{eqn:Threshold}); if the binary cross-entropy loss exceeds this threshold, the monitoring sample is classified as anomalous. Mathematically, we have, 
\begin{align}
   Loss(y,\hat{y})=-(y~log(\hat{y})+(1-y)~log(1-\hat{y})),
   \label{eqn:BCE loss}
\end{align}

\begin{align}
   if~~ L(y,\hat{y}) > {T_r},~ then ~Anomaly,
   \label{eqn:Threshold}
\end{align}
\\
We suggest the following stages for finding the ${T_r}$:
\begin{itemize}[leftmargin=*]
    \item  \textit{Step 1:} Calculate binary cross-entropy loss on a validation set of normal samples (Equation.~\ref{eqn:BCE loss}).
    \item  \textit{Step 2:} Sort the loss values and select a quantile for false positive tolerance (we defined 5\%).
    \item  \textit{Step 3:} Set the threshold at the corresponding loss value (Equation.~\ref{eqn:Threshold}, and see Fig.~\ref{fig:detection}).
    \item  \textit{Step 4:} Validate threshold performance on the validation set using evaluation metrics (we used F1-score).
\end{itemize}

It should be noted that the choice of threshold method depends on the dataset and anomaly characteristics. Experimentation and evaluation are crucial for determining the appropriate threshold value \cite{aguilar2022towards}.

\begin{table}[]
\centering
\caption{Deep auto-encoder CNN framework for detecting stealthy load-altering attacks, $N_x =\left | \mathcal{N} \right |/100 $}
\label{tab:CNN-AE}
\resizebox{\columnwidth}{!}{%
\begin{tabular}{ccccc}
\hline
\multicolumn{2}{c}{Operation Layer} & Number of Filters & Size of Each Filter & Size of Output Data \\ \hline
\multicolumn{2}{c}{Input Data}                       & -- & --         & $N_x \times 100 \times 3 $ \\ \hline
\multirow{2}{*}{Convolution Layer}   & Convolution   & 64 & $1 \times 3 $& $N_x \times 100 \times 64$ \\ \cline{2-5} 
                                     & ReLU          & -- & --         & $N_x \times 100 \times 64$ \\ \hline
Pooling Layer                        & Max pooling   & 1  & $1 \times 2$ & $N_x \times 50\times 64$   \\ \hline
Dropout Layer                        & Dropout (0.5) & 1  & --         & $N_x \times 50\times 64$   \\ \hline
\multirow{2}{*}{Convolutional Layer} & Convolutional & 32 & $1 \times 3$ & $N_x \times 50\times 32$   \\ \cline{2-5} 
                                     & ReLU          & -- & --         & $N_x \times 50\times 32$   \\ \hline
Pooling Layer                        & Max pooling   & 1  & $1 \times 3$ & $N_x \times 25\times 32$   \\ \hline
Dropout Layer                        & Dropout (0.5) & 1  & --         & $N_x \times 25\times 32$   \\ \hline
\multirow{2}{*}{Convolutional Layer} & Convolutional & 16 & $1 \times 3$ & $N_x \times 25\times 16$   \\ \cline{2-5} 
                                     & ReLU          & -- & --         & $N_x \times 25\times 16$   \\ \hline
\multirow{2}{*}{Convolutional Layer} & Convolutional & 32 & $1 \times 3$ &$ N_x \times 25\times 32$   \\ \cline{2-5} 
                                     & ReLU          & -- & --         & $N_x \times 25\times 32$   \\ \hline
Upsampling Layer                     & Upsampling    & 1  & $1 \times 2$ & $N_x \times 50\times 32$   \\ \hline
Dropout Layer                        & Dropout (0.5) & 1  & --         & $N_x \times 50\times 32$   \\ \hline
\multirow{2}{*}{Convolutional Layer} & Convolutional & 64 & $1 \times 3$ & $N_x \times 50\times 64$   \\ \cline{2-5} 
                                     & ReLU          & -- & --         & $N_x \times 50\times 64$   \\ \hline
Upsampling Layer                     & Upsampling    & 1  & $1 \times 2$ & $N_x \times 100\times 64$  \\ \hline
Dropout Layer                        & Dropout (0.5) & 1  & --         & $N_x \times 100\times 64$  \\ \hline
\multirow{2}{*}{Convolutional Layer} & Convolutional & 3  & $1 \times 3$ & $N_x \times 100\times 3$   \\ \cline{2-5} 
                                     & Sigmoid       & --  & --         &$ N_x \times 100\times 3$   \\ \hline
Output                               & --            & -- & --         & $N_x \times 100\times 3 $  \\ \hline
\end{tabular}%
}
\end{table}

\section{Numerical Results}\label{sec:numerical results}
In this section, we first assess the implications of CMAs on the DAM and RTM by examining the profits and penalties incurred by aggregators in the CAISO energy market \cite{jahangir2021novel} using the ACN-data dataset \cite{lee2019acn} for the EVCSs -- we defined a test case with 5000 EVCPs, constructed by combing the charging profiles of Caltech and JPL charging sites. Subsequently, we delve into the detection aspect of CMAs, utilizing the anomaly detection strategy introduced in Section~\ref{sec:detection scheme} (2D-CNN), along with three additional benchmark methods.
The simulations are conducted on a Windows PC with 11th Gen Intel(R) Core(TM) i7-1185G7 @ 3.00GHz processor, RAM: 16 GB.

\subsection{Impact on Energy Markets}\label{market:formulation}

\begin{figure}[]
\centering
\includegraphics[height=3 cm,width=9 cm,trim= 2 2 2 2,clip]{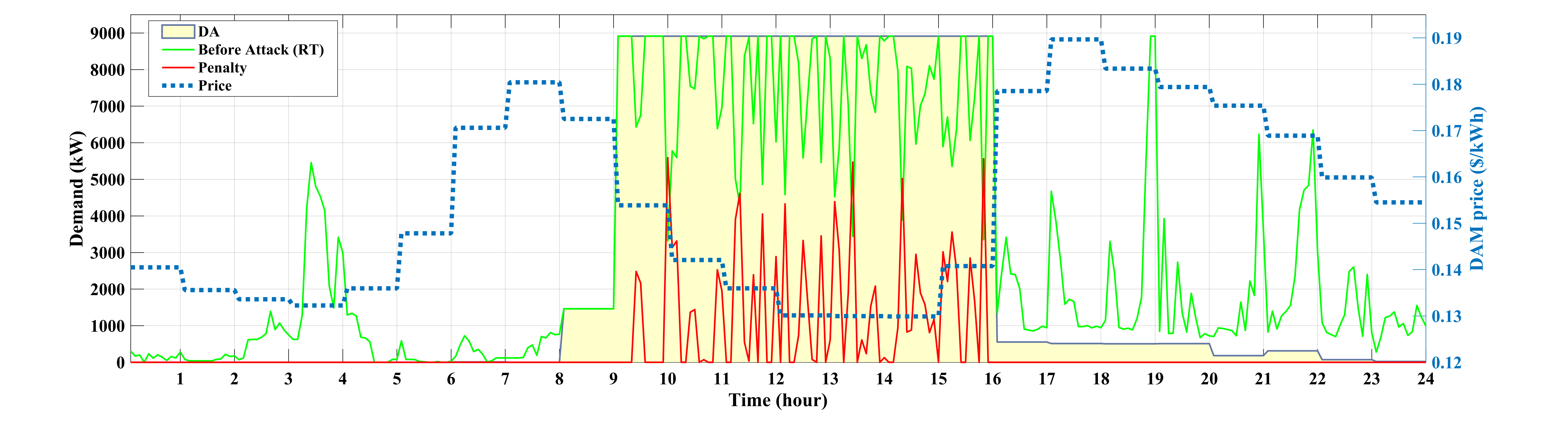}
\caption{DA bids, RT bids, and penalty (kW) before launching Type-1 CMAs}
\label{fig:energy_before}
\vspace{+0.2 cm}
\end{figure}

\begin{figure}[]
\centering
\includegraphics[height=3 cm,width=9 cm,trim= 2 2 2 2,clip]{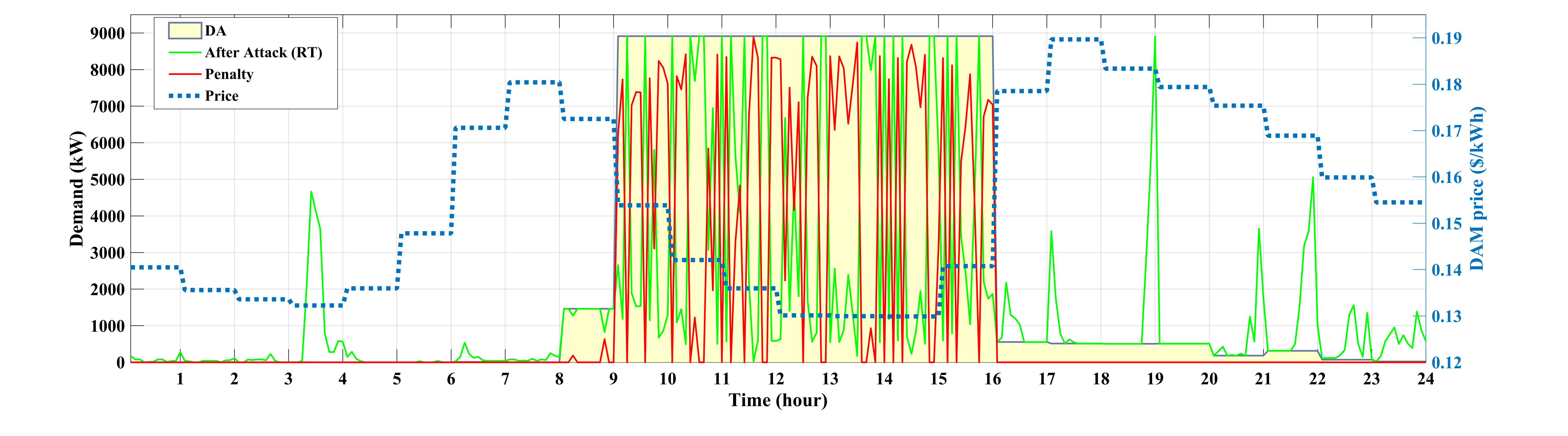}
\caption{DA bids, RT bids, and penalty (kW) after launching Type-1 CMAs}
\label{fig:energy_after}
\vspace{+0.2 cm}
\end{figure}

\begin{figure}[]
\centering
\includegraphics[height=4 cm,width=9 cm,trim= 2 2 2 2,clip]{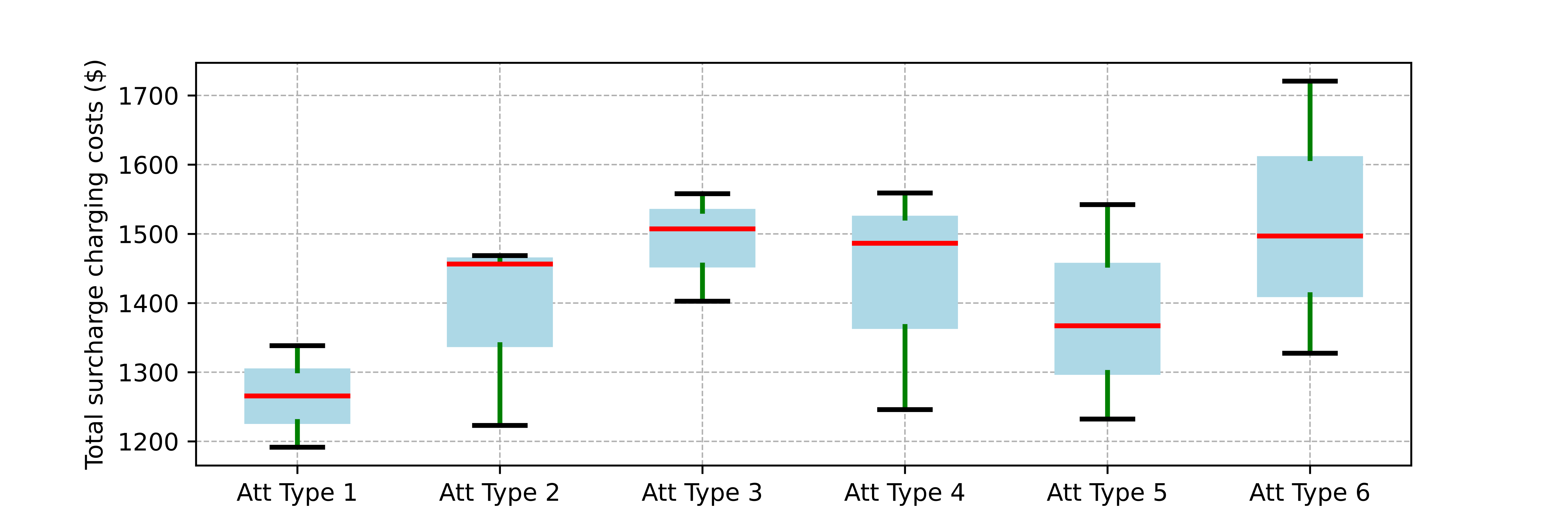}
\caption{Boxplot of Aggregated surcharge charging cost (\$) for different CMA scenarios}
\label{fig:attack impact energy market}
\vspace{+0.2 cm}
\end{figure}

To investigate the impact of CMAs on the charging costs of aggregators in the energy market, we consider a generalized replication of the CAISO energy market in our simulation results (Equations (3)-(5), full details are omitted for the sake of brevity, and can be found in \cite{jahangir2021novel}). Note that, our method does not mimic all the CAISO's real-time calculations of penalties and bid adjustments for market participants; this falls out of the scope of this research, complementary information is given in \cite{CAISO}. 
In the CAISO energy market, as explained in Section \ref{energy_market}, market participants must honor their awarded bids in the DAM on the following day; otherwise, they incur penalties and payment recessions in RTM. Additionally, they can adjust their bids on the following day in the RTM only by submitting incremental bids with RTM price. For the market data, we utilize the hourly DAM energy prices and five-minute RTM energy prices of CAISO node 0096WD\_7\_N001 on January 10, 2023. In this scenario, we make use of historical profiles of the ACN-data \cite{lee2019acn} and employ Monte Carlo (MC)\cite{jahangir2021novel} simulation to estimate the DA demand profile of EVCSs. We assume that the aggregator leverages this estimated value to participate in the DAM. Additionally, the aggregator takes charge of the bidding strategy outlined in Equations (3)-(5), aiming to minimize the overall charging cost while considering penalties and RTM incremental biddings. We explore two distinct scenarios: one without any CMAs, referred to as ``before attack" and another with CMAs, termed ``after attack" in Figs. \ref{fig:energy_before} and \ref{fig:energy_after}, as well as Table \ref{tab:energy market} -- to establish a fair comparison, the DA bids (generated by MC) will be held fixed in both scenarios. 
The simulation results reveal that upon executing Type-1 CMA, the aggregator experiences a substantial increase in the total penalty cost. Furthermore, the CMA limits the aggregator's flexibility in incremental bidding in the RTM, both factors leading to a notable 13\% increase in aggregated surcharge charging cost. 
These effects have significant implications for the aggregator's profit and trustworthiness within the energy market. In conclusion, Fig.~\ref{fig:attack impact energy market} illustrates the overall impact of all mentioned CMA types on the aggregated surcharge charging cost. The results highlight that the mixing CMA scenarios (Types 4-6) have a significantly higher impact compared to others (Types 1-3). This is due to the combination of shift and surge in demand experienced by the aggregator (as depicted in Fig.~\ref{fig:3D hist_all attack types}, and Fig.~\ref{fig:Charge_all attack types}) in mixing CMA scenarios (Types 4-6) that result in a significant disparity between the predicted DA and the actual demand experienced in RT.

\begin{table}[]
\label{tab:DA-Equations}
\scalebox{0.98}{
\renewcommand{\arraystretch}{1.3} 
\begin{tabular}{cc}
\hline
DAM Equations                                                             &    \\ \hline
${minimize~_{pch^{DA}_{n,tm,\hat{tm}}}}~~Cost^{DA}$                         & (3) \\
$Cost^{DA}=Cost^{CH,DA}+Cost^{EENC,DA}$                                   & (3a) \\
$Cost^{CH,DA}=\sum_{tm\in \tau }^{}{PCH_{tm}^{DA}\rho_{tm}^{DA}\Delta {tm}}$      & (3b) \\
$Cost^{EENC,DA}=\sum_{n\in \mathcal{N}}^{}{ENS_n^{DA}\rho_{ }^{EENC,DA}}$ & (3c) \\
For detailed formulation of ${pch^{DA}_{n,{tm},\hat{tm}}},{PCH_{tm}^{DA}}$\\ $Demand^{DA}_{n},$ and $ENS_n^{DA}$ please see \cite{jahangir2021novel}\\ \hline
RTM Equations                                                             &    \\ \hline
${minimize_{~ P_{{tm},\hat{tm}}^{INC,RT},~ P_{{tm},\hat{tm}}^{PEN,RT},~ pch_{n,{tm},\hat{tm}}^{RT} }}~~~Cost^{RT}$   & (4) \\
$Cost^{RT}=Cost^{INC,RT}+Cost^{PEN,RT}+Cost^{EENC,RT}$  & (4a) \\
$Cost^{INC,RT}=\sum_{{tm}\in \tau }\sum_{\hat{tm}\in \hat{\tau}}{P_{{tm},\hat{tm}}^{INC,RT}\rho_{tm,\hat{tm}}^{RT} \hat{\Delta {tm}}}$      & (4b) \\
$Cost^{PEN,RT}=\sum_{{tm}\in \tau }\sum_{\hat{tm}\in \hat{\tau}}{P_{{tm},\hat{tm}}^{PEN,RT}\rho^{PEN,RT} \hat{\Delta {tm}}}$ & (4c) \\
$Cost^{EENC,RT}=\sum_{n\in \mathcal{N}}{ENS_{n}^{RT}\rho^{EENC,RT} \hat{\Delta {tm}}}$ & (4d) \\
$PCH_{tm}^{DA}-P_{tm,\hat{tm}}^{PEN,RT}+P_{{tm},\hat{tm}}^{INC,RT}=PCH_{{tm},\hat{tm}}^{RT}$ & (4e) \\

For detailed formulation of ${pch^{RT}_{n,tm,\hat{tm}}},{PCH_{tm,\hat{tm}}^{RT}, ENS_{n}^{RT},}$ &\\ $P_{tm,\hat{tm}}^{INC,RT}$, $Demand^{RT}_{n}$ and $P_{tm,\hat{tm}}^{PEN,RT}$ please see \cite{jahangir2021novel}&\\ \hline
Total Cost &\\ 
$Cost^{Total}=Cost^{DA}+Cost^{RT}$ & (5)
\\ \hline
\end{tabular}
}
\end{table}

\begin{table}[]

\centering
\caption{Total charging cost of EVCSs in RT and DA markets before and after launching the Type-1 CMAs}
\label{tab:energy market}

\scalebox{0.9}{
\begin{tabular}{ccc}
\hline
\textbf{Different costs}       & \multicolumn{2}{c}{\textbf{Methods}}                             \\ \cline{2-3} 
\textbf{}                      & \textit{\textbf{Before Attack}} & \textit{\textbf{After Attack}} \\ \hline
DA EVs charging cost (\$)      & 9316                            & 9316                           \\
Penalty cost (\$)              & 1237                            & 4899                           \\
RT EV charging cost (\$)       & 2211                            & 670                            \\
RT EENC cost (\$)              & 416                             & 15                             \\
Total EVs charging cost   (\$) & 13182                           & 14903                          \\
Aggregated surcharge charging costs (\$)                     & --                              & 1720                           \\ \hline
\end{tabular}
}
\end{table}

\subsection{Attack Detection Results}

\begin{figure}[]
\centering
\includegraphics[height=6 cm,width=9 cm,trim= 2 2 2 2,clip]{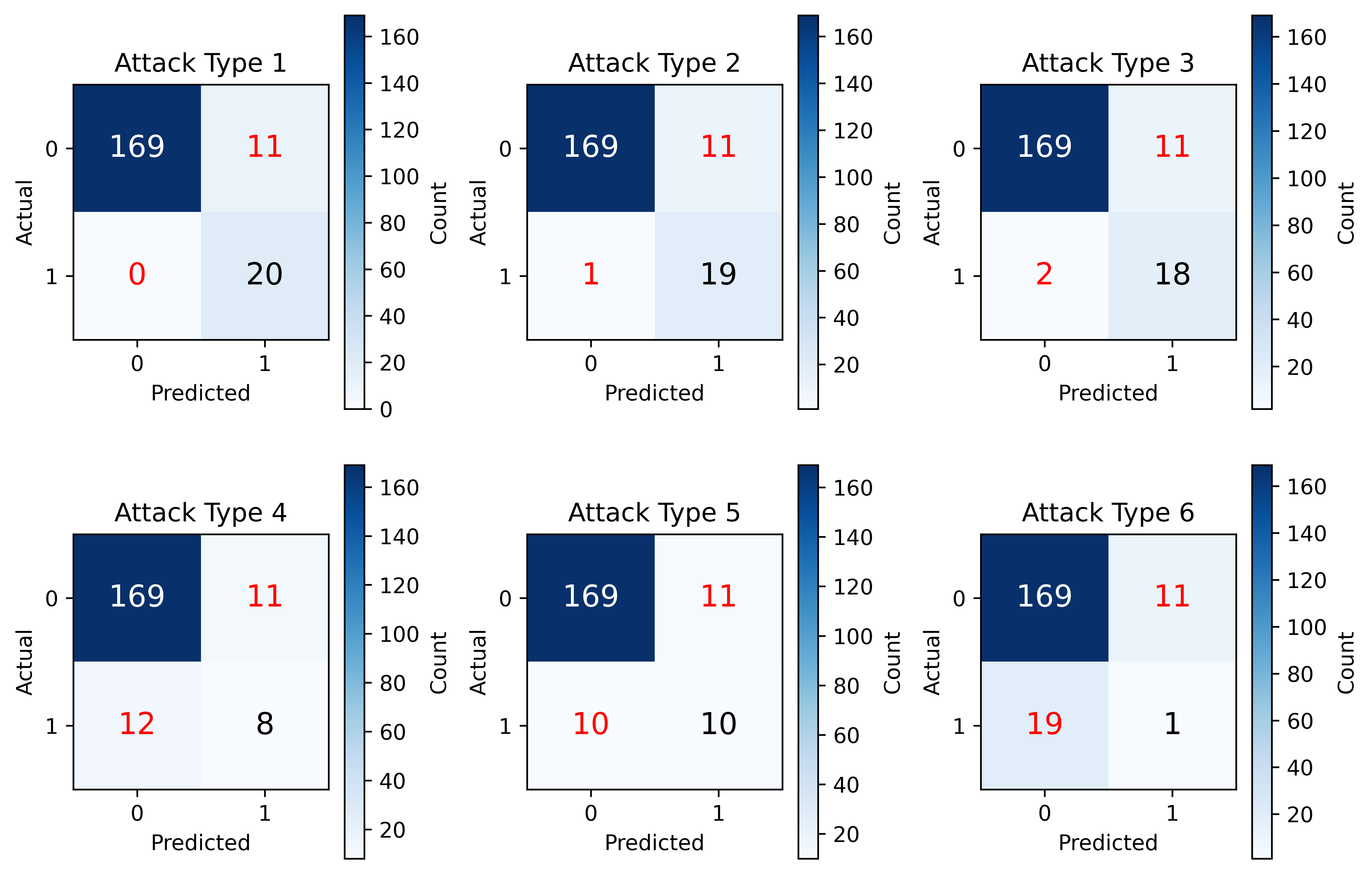}
\caption{
Detecting CMAs individually; each confusion matrix consists of 180 normal samples and 20 attack samples.}
\label{fig:detection_all}
\vspace{+0.2 cm}
\end{figure}

\begin{figure}[]
\centering
\includegraphics[height=4.5 cm,width=9 cm,trim= 2 2 2 2,clip]{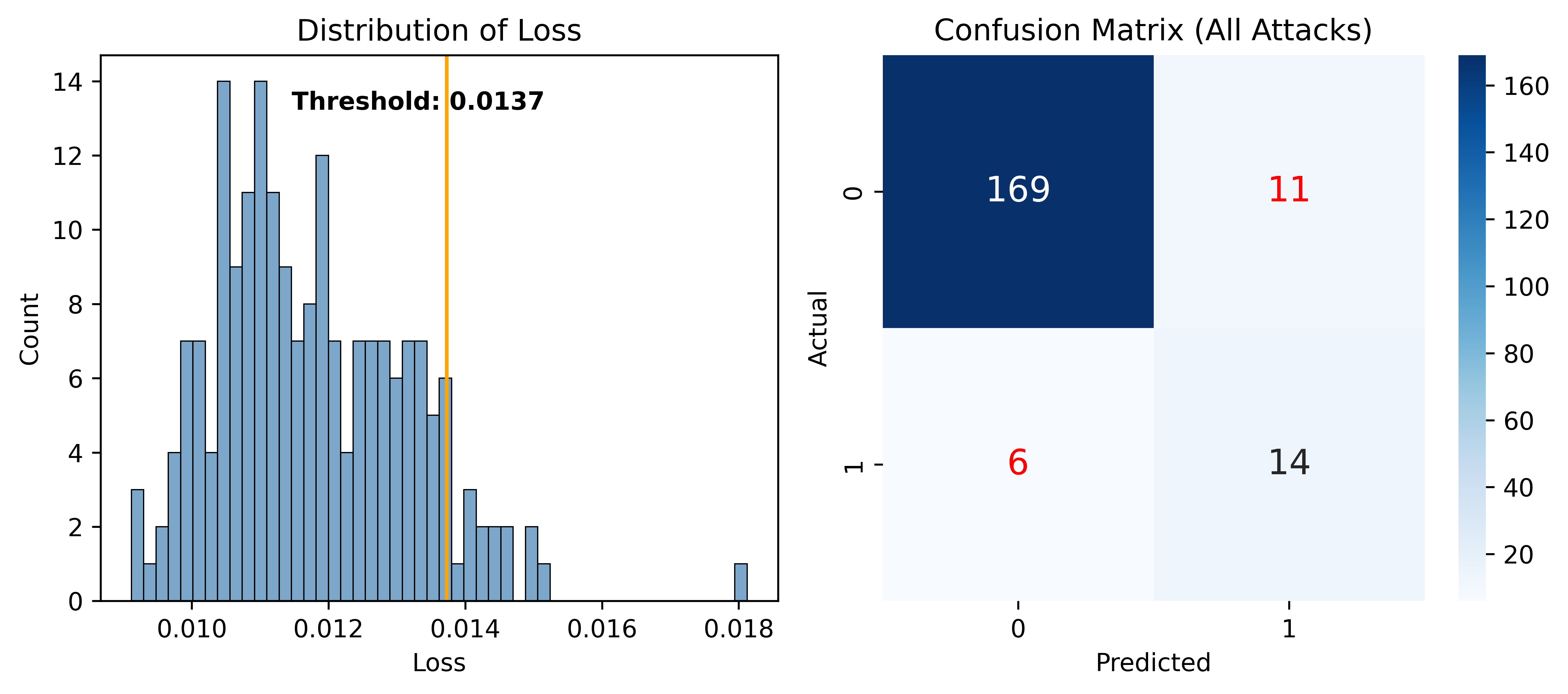}
\caption{
Detection of CMAs: (left) Finding the threshold value (${T_r}$) based on the mentioned strategy in Section~\ref{sec:detection scheme}, (right) confusion matrix with 180 normal samples and 20 attack samples (all types)}
\label{fig:detection}
\vspace{+0.2 cm}
\end{figure}

Next, we investigate the detection of CMAs using the anomaly detection approach suggested in Section \ref{sec:detection scheme}. Furthermore, we compare the performance with benchmark methods, including deep multilayer perception (MLP) auto-encoder, One-Class SVM (RBF Kernel), and Isolation Forest\cite{aguilar2022towards}. We create a test dataset consisting of 180 regular samples (90\%) and 20 attack samples (10\%) for each type of CMAs elaborated in Section \ref{launch_attack} (i.e., Types 1-6).
As shown in Fig.~\ref{fig:detection}, the threshold value for the anomaly detection method, discussed in Section \ref{sec:detection scheme}, has been determined as 0.0137. 



Table \ref{tab:detection results} presents results on the proposed method's effectiveness in detecting various types of CMAs as compared to benchmark anomaly detection strategies.
The results demonstrate that the proposed method outperforms other methods, most notably in accurately detecting the attacks and reducing false positives. In order to gain a deeper understanding of the stealthiness of different types of CMAs, we have provided the detection outcomes for each CMA scenario individually in Fig.~\ref{fig:detection_all}.
The findings demonstrate that Type-1 CMA, which solely considers the increases in the requested load demand, similar to the mentioned attacks in \cite{kabir2021two,wei2023cyber}, can be easily detected without any errors (0 False Positives (FP) out of 20 samples). Subsequently, Type-2 and 3 CMAs, which primarily involve manipulating the start time and end time respectively, exhibit a slightly higher level of stealthiness, resulting in 1 and 2 FP occurrences out of 20 samples respectively. 
In contrast, when it comes to Types-4, 5, and 6, which involve a combination of previous CMA scenarios, the situation in the detection tasks changes significantly. In these cases, the anomaly detection approach exhibits a notable increase in FPs, with more than 10 occurrences out of 20 samples. This observation highlights the stealthiness of these introduced scenarios compared to other CMA types -- using a blend of strategies to manipulate charging profiles creates minimal deviation from regular scenarios, potentially misleading unsupervised anomaly detection methods.

\begin{table}[]
\centering
\caption{Detection results of the proposed anomaly detection approach and other benchmark techniques}
\label{tab:detection results}
\scalebox{0.75}{
\begin{tabular}{ccccc}
\hline
Method                        & Precision & Recall & F1-score & Accuracy \\ \hline
2D CNN Auto-Encoder  (Table \ref{tab:CNN-AE})          & 0.88      & 0.91   & 0.89     & 91.24\%  \\
Deep MLP Auto-Encoder (256-128-64-128-256)           & 0.82      & 0.84   & 0.83     & 85.09\%  \\
One-Class SVM (RBF Kernel) \cite{aguilar2022towards}    & 0.81      & 0.90     &0.85     &89.17\%     \\
Isolation Forest   \cite{aguilar2022towards}         & 0.84       & 0.79      &  0.76   & 81.92\%     \\ \hline
\end{tabular}
}
\end{table}

\section{Conclusion}\label{conclusion}
This study has investigated the feasibility of deploying CMAs on EVCSs through the Open Charge Point Protocol (OCPP) perspective. Subsequently, a data-driven monitoring framework based on an unsupervised structure has been introduced, aimed at real-time detection of potential threats arising from manipulations in the charging settings. The numerical results demonstrate that the proposed CNN-based monitoring framework outperforms other benchmark machine learning anomaly detection methods, achieving over 5\% higher accuracy in detection results. Furthermore, this study has delved into the impact of CMAs from EVCSs on aggregators' profit and performance in the DA and RT energy markets, resulting in an approximately 13\% increase in the aggregated surcharge charging cost. The primary objective of this research has been to raise awareness among researchers and operators in the EV charging field about the potential risks posed by CMAs from EVCSs. It also emphasizes the need for developing sophisticated anomaly detection frameworks as modular solutions within smart charging platforms.





\bibliographystyle{IEEEtran}

\bibliography{ref}

\newpage

\end{document}

%% file: macros.tex
\newfont{\bbb}{msbm10 scaled 500}

\newfont{\bb}{msbm10 scaled 1100}

